\documentclass[12pt,epsfig]{article}
\usepackage{graphicx,amssymb}
\usepackage[figuresright]{rotating}
\newcommand{\noi}{\noindent}
\newcommand{\eq}{\begin{equation}}
\newcommand{\en}{\end{equation}}
\newcommand{\eqa}{\begin{eqnarray}}
\newcommand{\ena}{\end{eqnarray}}
\def\Journal#1#2#3#4{{#1}{\bf #2} (#4) #3}

\def\NPB{Nucl. Phys.~{\bf B}}
\def\NPPS{Nucl. Phys. Proc.~Suppl.~}
\def\PTPS{Prog.~Theor.~Phys.~Suppl.~}
\def\PLB{Phys.~Lett.~{\bf B}}
\def\PRL{Phys.~Rev. Lett.~}
\def\PRD{Phys.~Rev. ~{\bf D}}
\def\CMP{Comm.~Math.~Phys.~}
\def\ZPC{Z.~Phys.~{\bf C}}

\def\PRP{Phys.~Rept.~}
\def\PTP{Prog.~Theor.~Phys.~}
\def\RMP{Rev.~Mod.~Phys.~}
\def\JETP{JETP~Letters~}
\def\SJNP{Sov.~J.~Nucl.~Phys.~}
\def\JHEP{JHEP~}
\def\myre{{\rm Re}}

\begin{document}

\title{On the topological content \\ of $SU(2)$ gauge fields below $T_c$}
\author{
E.-M. Ilgenfritz$^a$, B. V. Martemyanov$^b$, M. M\"uller-Preussker$^c$, \\
S. Shcheredin$^c$ and A. I. Veselov$^b$ \\ $~$ \\
{\small
$^a$ Research Center for Nuclear Physics, Osaka University,
Osaka 567-0047, Japan} \\
{\small
$^b$ Institute for Theoretical and Experimental Physics,
Moscow 117259, Russia} \\
{\small
$^c$ Humboldt-Universit\"at zu Berlin, Institut f\"ur Physik,
D-10115 Berlin, Germany}
}

\maketitle

\vspace*{-12cm}
\begin{flushright}
{\small ITEP-LAT-2002-06 \\ HU-EP-02/22 \\ RCNP-Th02008} 
\end{flushright}
\vspace*{10cm}

\begin{abstract}
Finite temperature Euclidean $SU(2)$ lattice gauge fields generated in the 
confinement phase close to the deconfinement phase transition are subjected
to cooling. The aim is to identify long-living, almost-classical local 
excitations which carry (generically non-integer) topological charge.  
Two kinds of spatial boundary conditions (fixed holonomy and standard 
periodic boundary conditions) are applied.
For the lowest-action almost-classical configurations we find that their 
relative probability semi-quantitatively agrees for both types of boundary 
conditions. We find calorons with unit topological charge as well as
(anti-)selfdual lumps (BPS-monopoles or dyons) combined in pairs of 
non-integer (equal or opposite sign) topological charge.
For calorons and separated pairs of equal-sign dyons obtained by cooling 
we have found that (i) the gluon field is well-described by Kraan-van Baal 
solutions of the Euclidean Yang-Mills field equations and (ii) the lowest 
Wilson-fermion modes are well-described by analytic solutions of the 
corresponding Dirac equation.  
For metastable configurations found at higher action, the multi-center 
structure can be interpreted in terms of dyons and antidyons, using the 
gluonic and fermionic indicators as in the dyon-pair case. Additionally, 
the Abelian monopole structure and field strength correlators between 
the centers are useful to analyse the configurations in terms of 
dyonic constituents. 
We argue that a semi-classical approximation of the non-zero temperature 
path integral should be built on superpositions of solutions with 
non-trivial holonomy.
\end{abstract}

\section{Introduction}
\label{sect:introduction}
It is widely accepted that quark confinement in QCD is related to
some complex structure of the gauge field vacuum. A popular scenario views
the vacuum state as a dual superconductor~\cite{dualSC} in four dimensions 
which is unable to sustain strong gluon electric fields such that flux tubes
are formed due to a dual Meissner effect. The analogon to the Cooper pairs 
of usual superconductors are Abelian monopoles, in other words, condensed 
magnetic currents belonging to an Abelian subgroup of the strong gauge group.

The lattice evidence in support of this mechanism
rests either on the ability to localize the 
magnetic currents~\cite{dGT,lattice_evidence_percolation} 
as the worldlines of magnetic defects with respect to some gauge fixing 
prescription~\cite{AP}, or on the construction of a corresponding monopole 
creation operator~\cite{DG} in order to study its condensation.
The question how their condensation leads to confinement is relatively 
well-understood~\cite{DGL_1,DGL_2}, and while also the question of universality 
with respect to the gauge condition is partly answered 
affirmatively~\cite{universality_1,universality_2}, 
the reason {\it why} these monopoles are created 
in the QCD vacuum is far from being understood.

There is another working picture of the Yang-Mills 
vacuum~\cite{ILM_1,ILM_2,ILM_3,ILM_4} which has 
been applied very successfully to hadron physics, including basic features 
like chiral symmetry breaking and the $U_A(1)$ anomaly as well as 
details of spectroscopy~\cite{Rosner}. It is based on the instanton, 
a solution~\cite{BPST} 
of the Euclidean field equations. However, even in the form of the instanton 
liquid model~\cite{ILM_Rev}, when some local interaction between instantons is 
taken into account, instantons cannot be brought into relation with the 
third basic feature of QCD, confinement. There have been attempts to stretch 
the instanton liquid model to its limits in order to provide a confining 
string tension of sufficient strength between a static quark and 
antiquark~\cite{fukushima_1}, at least at intermediate distances of a few 
instanton radii. A recent assessment of the instanton-generated forces between
static charges has been discussed in Ref. \cite{fukushima_3}.

In lack of a satisfactory instanton-based mechanism of confinement,
many indications have been presented that instantons 
(or, more generally, carriers of topological charge) are closely related to the 
Abelian monopoles which are usually detected by gauge fixing procedures. 
This evidence stems from the observation made for one-instanton and 
many-instanton
configurations~\cite{Chernodub_Gubarev,Bornyakov_Schierholz} 
in the maximally Abelian gauge 
and from the observation of short range correlations between
topological charge and magnetic currents in genuine confining lattice 
configurations~\cite{Markum_Thurner,smoothing}.
On the other hand, also in artificial 
instanton liquid model ensembles (multi(anti)instanton configurations)
monopole network percolation has been observed~\cite{fukushima_2} as a 
prerequisite of monopole condensation. 
This has led to the conclusion that semi-classical, smooth gauge field 
configurations can give rise to networks of (light, condensed) magnetic 
monopoles even if these become discovered only as defects of the gauge 
fixing process. Thus, some reconciliation of the two pictures seems possible.

However, as long as the instantons are uncorrelated, the Abelian monopoles
and center vortices \cite{vortices} revealed by the respective gauge fixing 
are quantitatively insufficient to provide confinement~\cite{fukushima_3}. 
Moreover, the approximate Casimir scaling at intermediate distance that 
the strong force acting on charges of different representations obeys
is also strongly broken in the instanton 
liquid ensemble~\cite{fukushima_3}. This makes it less likely 
that instantons are related to confinement even 
if some important correlation would be included into the model.

Nowadays, in a fresh attempt to relate instanton physics to the confinement
property of the Yang-Mills vacuum, instantons dissociated into meron pairs 
have been investigated with respect to their monopole and center vortex 
content~\cite{MN}. Our present work, which also aims to extend the instanton 
picture, starts from a somewhat different corner, finite temperature.
All the criticism with respect to an instanton mechanism of confinement
applies also to finite temperature Yang-Mills theory below the deconfinement 
temperature, where periodic instantons (calorons) are the ingredients of the 
instanton liquid model~\cite{GPY}. More precisely, there are periodic 
instantons {\it with trivial holonomy} which are thought of being the
bricks of topological structure. An isolated caloron with trivial holonomy 
is a periodic classical solution of the Euclidean equations of motion with
trivial asymptotics of the Polyakov loop at infinity, 
$P(\vec{x}\to\infty) = z \in Z(N)$. These are the ''old'' calorons first 
considered by Harrington and Shepard~\cite{HS}.  

During the last years new caloron solutions have been found and studied 
by Kraan and van Baal (KvB)~\cite{KvB,big_paper} 
which correspond to a non-trivial 
asymptotic holonomy $P(\vec{x}\to\infty) ~{\not \in}~Z(N)$. 
From our point of view, a particularly interesting feature is that, in the 
generic case, these configurations are composites of BPS-monopoles~\cite{BPS}
(or dyons) {\it i. e.} lumps carrying non-integer topological charge. 
A semi-classical description of the (finite temperature) vacuum in the
confinement phase, based on these new calorons still remains to 
be developed. Some new perspectives have been pointed out in 
Ref. \cite{GPGAMvB}.
In the present paper we investigate metastable configurations which 
appear in the process of cooling~\cite{ILMPSS} in the region of few times the 
instanton action. These are excitations which might become the building blocks 
of such a semi-classical description.
The topological content, which appears at this deep, almost classical 
level of cooling, is itself characteristic for the phase (confining or 
deconfined) where the configurations are taken from before they are cooled. 
Still, we do not claim that the number and size of topological lumps 
obtained in this way immediately characterizes the finite-temperature 
vacuum.
  
The outcome of cooling with normal periodic boundary conditions has been
compared with that of cooling with spatial boundary conditions of prescribed
holonomy, which is enforced by cold boundary links appropriately chosen.
As for the confinement phase, the dependence on the boundary conditions
is weak. A closer check shows that, after cooling with periodic boundary
conditions, the average Polyakov loop over regions of low action and 
topological density is not restricted to trivial holonomy. In other
words, in the confined phase non-trivial holonomy boundary conditions for 
topological charge lumps are provided by the dynamics itself. They do not 
need to be enforced. In the confined phase these are the natural fluctuating 
asymptotic conditions for the semi-classical excitations.

Similarly to the studies of the monopole and vortex content of meron~\cite{MN} 
pairs, we study here also the monopole content
and other signatures of the simplest of these configurations (to be interpreted 
as dyon-dyon and dyon-antidyon pairs) as potential semi-classical building 
blocks of a confining vacuum. Preliminary results have been presented
in~\cite{IMPV,IMMPV1,IMMPV2}.
  
In contrast to the meron studies~\cite{MN}, instead of tayloring particular
semi-classical configurations, in the present work we are going 
to demonstrate the emergence under cooling of metastable lattice 
configurations with low action which (distinguished by action plateaus) 
resemble the new calorons (dyon-dyon pairs). 
In this case, where an analytical solution is available, we show that
action density, topological density and the profile of the fermionic zero modes 
on the lattice can be fit simultaneously with the corresponding expressions 
for the new caloron KvB solution~\cite{KvB,CKvB}.
It is worth noticing that at actions around one instanton action also 
configurations pop up under cooling which resemble dyon-antidyon pairs. 
This interpretation is suggested by action and 
topological density and is corroborated by the zero modes of the Wilson fermion 
matrix and the Polyakov loop profile. Lacking an analytical solution for these 
configurations of mixed (anti)selfduality it is impossible to present a fit of
the various distributions.

Also for higher metastable plateaus with an action of few times the instanton 
action a multi-center structure of few-lump dyon-antidyon mixed configurations 
can be recognized having a broad distribution of holonomy away from the lumps 
of action and topological charge. We do not claim that this directly proves 
that the finite temperature vacuum is composed exactly as that gas of 
dyons and antidyons. 
Nevertheless, it is remarkable that cooling starting from genuine Monte Carlo 
configurations in the {\it deconfined} phase ends in completely different 
low-action configurations as it has been already claimed in an early 
related work by Laursen and Schierholz \cite{LS}.

The paper is organized as follows.

In Section \ref{sect:formulae} we will collect some necessary formulae covering
the zero-temperature instanton, the well-known selfdual finite temperature 
instanton (the ''old caloron''), 
the new KvB calorons, monopoles as instanton 
chains etc. This will be needed for the fit of dyon-dyon pairs becoming
visible on the lattice.

Section \ref{sect:detecting} provides all necessary lattice definitions, 
in particular the boundary conditions and the observables considered in 
order to identify KvB and other solutions.

In Section \ref{sect:simple_lumps} we describe the results of a first part 
of our simulations where equilibrium configurations are cooled down to an 
action near to the one-instanton action. Although we cool to this low level 
of action, we find the topological content (and the dependence on the boundary 
conditions) very different, depending on whether the original Monte Carlo 
ensemble was describing the confinement or deconfinement phase.

In Section \ref{sect:gases} we report on a second part of our study, 
where higher lying action plateaus, appearing in the process of cooling with 
periodic boundary conditions, are identified with respect to their dyon content.
  
Section \ref{sect:conclusions} presents the conclusions. 

\section{Calorons with trivial and non-trivial holonomy
in continuum $SU(2)$ Yang-Mills theory}
\label{sect:formulae}
Throughout this paper $SU(2)$ gauge theory in four-dimensional Euclidean
space is considered. 
We start the description of caloron solutions with generically 
non-trivial holonomy
with the well-known one-instanton solution \cite{BPST,tH}.
In the singular gauge its gauge potentials are given by
\eq  \label{eq:inst}
A_\mu(x)=\frac{\tau_a}{2}\bar{\eta}^a_{\mu\nu}\partial_\nu\log\phi(x)\,,
~~~~\phi(x)=1 + \frac{\rho^2}{(x-x_0)^2}\,, \qquad \mu, \nu = 1, \cdots, 4
\en
where $\bar{\eta}^a_{\mu\nu}$ is the 't Hooft tensor, $\tau_a/2$ -
denote the generators of the $SU(2)$ group, $\rho,x_0$ are the size of the
instanton and the $4D$ position of its center. 
For further use we introduce the following notations
\eqa  \label{eq:notation}
\phi(x) &=& \frac{\psi(x)}{\hat{\psi}(x)}\,, \nonumber \\ 
\psi(x) &=& 2\pi^2((x-x_0)^2+\rho^2)\,, ~~~~\hat{\psi}(x) = 2\pi^2(x-x_0)^2\,,
\\
\chi(x) &=& 1-\frac{1}{\phi}=\frac{2\pi^2\rho^2}{\psi}  \nonumber
\ena
and rewrite the instanton fields in the form
\eqa \label{eq:instnew}
A_\mu(x) = &&\frac{1}{2}\bar{\eta}^3_{\mu\nu}\tau_3\partial_\nu\log\phi(x)
\nonumber \\
&+& \frac{1}{2}\phi(x)\myre\left((\bar{\eta}^1_{\mu\nu}-i\bar{\eta}^2_{\mu\nu})
(\tau_1+i\tau_2)\partial_\nu\chi(x)\right)\,.
\ena
The next step is to construct the time-periodic instanton or
caloron \cite{HS}. For this purpose we generalize the function $\phi$
in Eq. (\ref{eq:inst})
\eqa \label{eq:calor}
\phi(x) &=& 1 + \sum_{n=-\infty}^{n=\infty}
\frac{\rho^2}{((\vec{x} - \vec{x}_0)^2+(t-nb)^2)} \nonumber \\
&=& 1+\frac{\pi\rho^2}{br}
\frac{\sinh(\frac{2\pi r}{b})}{\cosh(\frac{2\pi r}{b})-\cos(\frac{2\pi t}{b})}
\,,
\ena
where $r=|\vec{x}-\vec{x}^{(0)}|$ and $t=x_4-x_4^{(0)}$.
$b$ is the time period of the periodic instanton solution.
For simplicity, $b=1$ is assumed in the following. Physically, the periodicity 
$b$ has to be identified with the inverse temperature $T^{-1}$.
Then the caloron potentials are given by Eq. (\ref{eq:instnew}) with
\eqa \label{eq:notationnew}
\phi(x) &=& \frac{\psi(x)}{\hat{\psi}(x)}\,, \nonumber \\
\psi(x) &=& \cosh(2\pi r)-\cos(2\pi t)+\frac{\pi\rho^2}{r}\sinh({2\pi r})\,,
~~~~\hat{\psi}(x) = \cosh(2\pi r)-\cos({2\pi t})\,, \\
\chi(x) &=& 1-\frac{1}{\phi}=\frac{\pi\rho^2\sinh({2\pi r})}{\psi r}\,.\nonumber
\ena
This solution can be viewed as a chain of instantons aligned along the temporal
direction, separated by distance $b$ but 
with the same orientation in color
space. In the limit $\rho \rightarrow \infty$ it 
can be transformed into a static solution by  
an appropriate gauge transformation. This static
object can be identified with a BPS monopole \cite{BPS} in Euclidean space,
where the fourth component $A_4$ plays the r\^ {o}le of the Higgs field
in the adjoint representation. The solution has both electric and magnetic
charge. Therefore, we call it also {\it dyon} ($D$).

Constructing a caloron with nontrivial holonomy \cite{KvB} is the next step.
It can be approximately viewed as an instanton chain with periodicity $b$, too, 
where each of the instantons is rotated with respect to the previous 
one by an angle $4\pi\omega$ in color space, $\omega$ being the parameter of 
holonomy as we will see
below. The rotation axis can be any, for definiteness let 
us take the third one. The words {\it approximately viewed} mean that when the 
instantons are well separated ($\rho/b << 1$) the fields near the instanton 
centers look approximately as described above. The caloron field with nontrivial
holonomy is described again by Eq. (\ref{eq:instnew}) but now with \cite{KvB}
\eqa  \label{eq:chi}
\phi(x) &=& \frac{\psi(x)}{\hat{\psi}(x)}\,, \nonumber \\
\psi(x) &=& -\cos(2\pi t)+\cosh(4\pi r\bar{\omega})\cosh(4\pi s \omega)
 + \frac{(r^2\!+\!s^2\!+\!\pi^2\rho^4)}{2rs}
   \sinh(4\pi r \bar{\omega}) \sinh(4\pi s\omega)  \nonumber \\
 && + \pi\rho^2 \left(s^{-1}\sinh(4\pi s\omega)\cosh(4\pi r\bar{\omega})
 + r^{-1}\sinh(4\pi r\bar{\omega})\cosh(4\pi s\omega)\right), \\
\hat{\psi}(x) &=& -\cos(2\pi t)+\cosh(4\pi r\bar{\omega})\cosh(4\pi s\omega)
 + \frac{(r^2 \!+\!s^2\!-\!\pi^2\rho^4)}{2rs}\sinh(4\pi r\bar{\omega})
   \sinh(4\pi s\omega)\,, \nonumber \\
\chi(x)&=& e^{4\pi it\omega}\frac{\pi\rho^2}{\psi}
  \left\{e^{-2\pi it}s^{-1}\sinh(4\pi s\omega)+r^{-1}\sinh(4\pi r\bar{\omega})
   \right\} \nonumber
\ena
instead of Eq. (\ref{eq:notationnew}). 
The holonomy parameters $\omega$ and $\bar{\omega}$ 
are related to each other as
$\bar{\omega} = 1/2 - \omega, ~~0 \le \omega \le 1/2$.
$r=|\vec{x}-\vec{x}_1|$ and $s=|\vec{x}-\vec{x}_2|$ 
are the $3D$ distances to the locations of the two centers of the new 
caloron solution,
$\pi\rho^2/b = d \equiv |\vec{x}_1-\vec{x}_2|$ is expressed by 
the distance between the centers. 
From Eqs. (\ref {eq:chi}) it can be easily seen
that, when $\omega \rightarrow 0$ or $\bar{\omega} \rightarrow 0$, the caloron
with nontrivial holonomy turns into the Harrington-Shepard
caloron described by Eqs. (\ref{eq:notationnew}). When the separation
between the centers becomes large, $d = \pi\rho^2/b >> 1$, 
two well-separated constituents emerge which are static in time. 
The ''mass'' ratio of these dissociated constituents is equal
to $\bar{\omega}/\omega$. Since the full solution is selfdual,
the ratio is the same for the action as for the equal-sign topological charge
carried by the constituents, the latter summing up to one unit of 
topological charge, $Q_t = 1.$
The separated constituents form a pair of BPS monopoles (or dyons) with
opposite magnetic charges. In the following we 
will call it a $DD$ pair,
while non-dissociated calorons will be denoted as $CAL$, 
although all these objects represent limiting cases of one and 
the same solution.
The single dyon originally obtained from the Harrington-Shepard caloron in the 
infinite size limit $\rho \rightarrow \infty$ can be recovered from the new
solution by sending the mass of the second constituent to zero and 
simultaneously its position to spatial infinity.

The action density in all three cases described above can be expressed by
a simple formula
\eq \label{eq:solutionact}
 s(x)=-\frac{1}{2}\partial^2_\mu\partial^2_\nu \log \psi(x)~.
\en

So far, the new caloron solution is in the so-called algebraic 
gauge~\cite{big_paper}. 
It can be made periodic by a gauge transformation 
which is non-periodic in time $g(x)=e^{-2\pi it\omega\cdot\tau_3}$,
\eqa \label{eq:solutionpg}
A_\mu^{\rm per} = && \frac{1}{2}\bar{\eta}^3_{\mu\nu}\tau_3\partial_\nu\log \phi
\\
&+& \frac{1}{2}~\phi~\myre\left((\bar{\eta}^1_{\mu\nu}-i\bar{\eta}^2_{\mu\nu})
(\tau_1+i\tau_2)(\partial_\nu+4\pi i\omega\delta_{\nu,4})\tilde\chi\right)
+\delta_{\mu,4}~2\pi \omega\tau_3\,, \nonumber
\ena
where
\eq
\tilde\chi \equiv e^{-4\pi it\omega}\chi
   = \frac{\pi\rho^2}{\psi}\left\{e^{-2\pi it}s^{-1}\sinh(4\pi s\omega)
                                  + r^{-1}\sinh(4\pi r\bar{\omega})\right\}.
\en
Now the time component of the caloron potential has become nonzero at
spatial infinity. We can define the {\it holonomy} which becomes
non-trivial
\eq \label{eq:polloop}
{\cal P}(\vec{x}) =
\mathrm{P}~\exp~\left(~i~\int_0^{b}A_4(\vec{x},t)dt~\right) \rightarrow
{\cal P}_{\infty} =  e^{2\pi i\omega\tau_3}
\qquad \mbox{for} ~|\vec{x}| \to \infty\,.
\en
In terms of $\omega$, the normalized trace of the holonomy, 
the Polyakov loop which we shall take as a direct 
{\it measure of the holonomy}, at spatial infinity 
becomes 
\eq
L(\vec{x}) \equiv \frac{1}{2}\mathrm{tr}~{\cal P}(\vec{x}) \rightarrow
L_{\infty} \equiv \frac{1}{2}\mathrm{tr}~{\cal P}_{\infty} = \cos(2\pi\omega)\,.
\en
In the special case $\omega = \bar{\omega} = 1/4$ the measure of holonomy
is equal to zero and the two constituent dyons 
acquire equal mass, {\it i. e.} equal action and topological charge.
For our later purposes it is  
useful to remember that the Polyakov loop has peaks $L(\vec{x}) = \pm 1$
very close to the center positions $\vec{x}=\vec{x}_{1,2}$ of the
constituents \cite{KvB}.

For well-separated dyons, when the functions $\phi(x)$ and $\psi(x)$ are almost 
time independent, the strongest 
time dependence comes from the first part of the function 
$\tilde\chi(x)$. This dependence is represented by the phase $e^{-2\pi it}$ 
and is nothing else but the homogeneous rotation of the first dyon,  
which has $L(\vec{x_1}) = - 1$, in color space around the third axis with angle 
$2\pi$ over the period. The second dyon with $L(\vec{x_2}) = + 1$ is static. 
Such a relative rotation of two dyons (that form a monopole-antimonopole pair) 
gives the so-called {\it Taubes winding} necessary two produce unit 
topological charge from a monopole-antimonopole pair \cite{Taub}.
One can detect this rotation in a gauge invariant fashion by looking at the
gauge invariant field strength correlator defined on each constant-time slice
and watching its evolution over the periodicity interval $b$. 
The field strength is selfdual ($E^a_k = B^a_k$ or antiselfdual, 
$E^a_k = - B^a_k$) everywhere in 
the KvB caloron. The electric field $E^a_k(\vec{x}_i,t)$ at 
the centers of both dyons $i=1,2$ is proportional to an orthogonal matrix 
(in both the $SU(2)$ color and space indices).
Thus the three ($k=1,2,3$) components $E^a_k(\vec{x}_1)$ 
of the electric field form vectors in color space which represent a 
local reference frame at the center of the first dyon. 
The comparison with the local frame at 
the center of the other dyon, $\vec{x}_2$, can be made in a gauge 
invariant manner by connecting the centers 
by the (fixed time) Schwinger line parallel transporter
\eq \label{eq:Schwinger}
{\cal S}(\vec{x}_1,t;\vec{x}_2,t)=
\mathrm{P}~\exp~\left(~i~\int_{\vec{x}_2}^{\vec{x}_1} 
A^a_k(\vec{x^{\prime}},t)\frac{\tau^a}{2}~dx^{\prime}_k~\right) \, . 
\en
Using this one can form the gauge invariant field strength product
\eq \label{eq:Rotation}
R^{12}_{kl}(t) = 
\mathrm{tr} \left(~E_k(\vec{x}_1,t)~S    (\vec{x}_1,t;\vec{x}_2,t)~
                   E_l(\vec{x}_2,t)~S^{+}(\vec{x}_1,t;\vec{x}_2,t)~\right)
\en
which is again an orthogonal matrix. This matrix performs a full rotation 
with the Euclidean time $t$ running from $0$ to $b$. 

Finally, let us comment on the zero-mode eigenfunctions of the fermionic
massless Dirac operator in the background of the KvB solutions. 
They have been studied analytically in \cite{GPGAPvB} and \cite{CKvB}. 
One finds closed solutions depending on the type of (anti)periodic 
boundary conditions (b.c.) 
imposed on the fermion fields in the imaginary time direction.
In case of well-separated dyon pairs, {\it i. e.} for $d = \pi\rho^2/b >> 1$,
the zero eigenmode densities become very simple expressions
\eqa \label{eq:fermiondensity}
|\psi^{-}(x)|^2 &=& - \frac{1}{4 \pi} \partial_{\mu}^2 
                    \left[ \tanh(2 \pi r \bar{\omega}) / r \right] \qquad
                    \mbox{for antiperiodic b.c.}\,, \\
|\psi^{+}(x)|^2 &=& - \frac{1}{4 \pi} \partial_{\mu}^2 
                    \left[ \tanh(2 \pi s \omega)       / s \right] \qquad
                    \mbox{for periodic b.c.}\,. \nonumber
\ena
This means that the zero-mode eigenfunctions are localized always around one
of the constituents of the KvB solution, for antiperiodic b. c. at that 
constituent which has $L(\vec{x})= - 1$ at its center $\vec{x}_1$. 
Switching to periodic b.c. for the fermion fields the zero mode localization
jumps to the other constituent monopole of the gauge field. Therefore,
the fermionic zero modes provide a convenient way to identify a monopole-pair 
structure in the gauge fields.   

\section{Detecting dyons and calorons on the lattice}
\label{sect:detecting}
Our first aim was to detect the simplest dyon configurations in the context 
of finite temperature lattice simulations. For this purpose we have 
considered $SU(2)$ lattice gauge theory on an asymmetric lattice using the 
standard Wilson plaquette action with coupling $\beta=4/g_0^2$,
\eqa \label{eq:action}
S = \sum_{\vec{x},t} s(\vec{x},t) &=& \sum_{\vec{x},t} \sum_{ \mu < \nu}
                    s(\vec{x},t;\mu,\nu), \\
s(\vec{x},t;\mu,\nu) &=& \beta~(1 - \frac{1}{2}\mathrm{tr}~U_{x,\mu\nu}),\qquad
               U_{x,\mu\nu} = U_{x,\mu} U_{x+\hat{\mu},\nu}
               U^{\dagger}_{x+\hat{\nu},\mu} U^{\dagger}_{x,\nu} \nonumber
\ena
and periodic boundary conditions in all four (toroidal) directions.
For simplicity the lattice spacing is set to $a=1$.
The lattice size will be $N_s^3 \times N_t$ with the spatial extension 
$N_s = 16$ or $24$ and with $T^{-1} \equiv N_t = 4$. 
For $N_t=4$ the model is known to undergo 
the deconfinement phase transition at
the critical coupling $\beta_c \simeq 2.299$ \cite{Bielefeld}. 
Throughout this paper we 
concentrate on the confinement phase, i.e. $\beta \le \beta_c.$

We shall generate the quantum gauge field ensemble $\{ U_{x,\mu} \}$ by
simulating the canonical partition function 
using the standard heat bath Monte Carlo method. 
The equilibrium field configurations will be cooled by iteratively minimizing 
the action $S$. Usually, cooling in one or another form is used
in order to smooth out short-range fluctuations, while 
(initially) leaving some long-range physics intact. 
The cooling method applied here is the standard relaxation 
method described long time ago in \cite{ILMPSS}. 

This method, if applied without any further limitation, easily finds
approximate solutions of the lattice field equations as shoulders 
(plateaus) of action in the relaxation history. Under certain circumstances, 
this defines and preserves the total topological charge of a configuration.
However, the short-range structure of the vacuum fields is changed. 
Still, the type of classical solutions, which are being
selected, depends on the phase which the quantum ensemble $\{ U_{x,\mu} \}$ 
is meant to describe. 
We want to investigate smoothed fields at different stages of cooling, 
by using a stopping criterium which selects the plateaus in a
given interval of action.
First, in Sect. 4, we focus at the lowest action plateaus, i.e.
$ S = (0.5 \cdots 1.5)~S_{\mathrm{inst}} $, where 
$S_{\mathrm{inst}} = 8 \pi^2 / g_0^2$
denotes the action of a continuum instanton. Lateron, in Sect. 5, we shall 
describe more complex approximate solutions found by stopping
at higher plateaus.  

The smoothed fields will be analyzed according to 
the spatial distributions of the following observables:
\begin{itemize}
\item {\it action density} computed from the local plaquette values and
  averaged with respect to the time variable:
  \eq \label{eq:actdensity}
  \varsigma(\vec{x}) = \frac{1}{N_t} \sum_t s(\vec{x},t) \,;
  \en
\item {\it topological density} computed with the standard discretization
and averaged over the time variable:
  \eq \label{eq:topdensity}
  q_t(\vec{x}) = - \frac{1}{N_t} \frac{1}{2^4 \cdot 32 \pi^2} 
               \sum_t\left( \sum_{\mu,\nu,\rho,\sigma=\pm 1}^{\pm 4} 
           \epsilon_{\mu\nu\rho\sigma}
           \mathrm{tr} \left[ U_{x,\mu\nu} U_{x,\rho\sigma} \right]
           \right) \,;
  \en
\item {\it Polyakov loop} defined as:
  \eq \label{eq:latpolloop}
  L(\vec{x}) = \frac{1}{2} \mathrm{tr} \prod_{t=1}^{N_t} U_{\vec{x},t,4} \,,
  \en
  where the $U_{\vec{x},t,4}$ represent the
  links in time direction;
\item {\it Abelian magnetic fluxes and monopole charges} defined within the
  maximally Abelian gauge (MAG). The latter is found by maximizing the 
  gauge functional
  \eq \label{eq:mag}
  F[g] = \sum_{x,\mu}
       \mathrm{tr} ( U^{g}_{x,\mu} \tau_3 U^{g \dagger}_{x,\mu} \tau_3 )\,,
  \en
  with respect to gauge transformations $U_{x,\mu} \rightarrow U^g_{x,\mu}
  = g(x) U_{x,\mu} g^{\dagger}(x + \hat{\mu})$. Abelian
  link angles $\theta_{x,\mu}$ are then defined by Abelian projection onto
  the diagonal $U(1)$ part of the link variables $U_{x,\mu} \in SU(2)$.
  According to the DeGrand-Toussaint prescription \cite{dGT} 
  a gauge invariant magnetic flux $\bar{\Theta}_p$ through an oriented 
  plaquette $p \equiv (x,\mu\nu)$ 
  is defined by splitting the plaquette
  $\Theta_p = \theta_{x,\mu} +\theta_{x+\hat{\mu},\nu}
              -\theta_{x+\hat{\nu},\mu}-\theta_{x,\nu}$ 
  into 
  $\bar{\Theta}_p = \Theta_p + 2\pi n_p, ~~n_p=0,\pm 1,\pm 2$ such that
  $\bar{\Theta}_p \in (-\pi, +\pi]$. 
  The magnetic charge of an
  elementary 3-cube $c$ is then
  $m_c = \frac{1}{2\pi} \sum_{p \in \partial c} \bar{\Theta}_p\,;$ 
\item {\it eigenvalues and eigenmode densities} of the non-Hermitean 
  standard Wilson-Dirac operator
  \eq \label{eq:fermionoperator}
  \sum_{y,s,\beta} D[U]_{xr\alpha,ys\beta} ~\psi_{s\beta}(y) = 
                             \lambda ~\psi_{r\alpha}(x) 
  \en
  with
  \eqa \nonumber
  D[U]_{xr\alpha,ys\beta} &=& 
          \delta_{xy} \delta_{rs} \delta_{\alpha\beta} \\
          &-& \kappa \sum_{\mu} 
   \left\{ \delta_{x+\hat{\mu},y}\, 
          ({\bf 1}_D - \gamma^{\mu} )_{rs} 
          (U_{x,\mu})_{\alpha\beta} 
        + \delta_{y+\hat{\mu},x}\, 
          ({\bf 1}_D + \gamma^{\mu} )_{rs} 
          (U^{\dagger}_{y,\mu})_{\alpha\beta} \right\} \nonumber
  \ena
  studied both with time-antiperiodic and time-periodic boundary conditions.
  For our purposes it will be sufficient to consider this operator which 
  breaks explicitely chiral invariance. To use a chirally improved lattice 
  Dirac operator would be a next step. We find the $\lambda$ spectrum
  and the eigenfunctions with the help of the implicitely restarted
  Arnoldi method \cite{neff,arnoldi} and use the standard ARPACK code package
  for this aim.  
\end{itemize}

\noi
For production and cooling of the equilibrium gauge field configurations
we shall use two kinds of spatial boundary conditions
as in \cite{IMMPV1} and with preliminary results presented in \cite{IMMPV2}
\begin{enumerate}
\item standard periodic boundary conditions (p.b.c) on the $4D$ torus;
\item fixed holonomy boundary conditions (f.h.b.c.): fixed holonomy is 
realized by cold timelike links $U_{\vec{x},t,4}$ on the spatial boundary 
$\Omega$. 
\end{enumerate}

For clarity we stress, 
that the second case is periodic, too, but for the spatial boundary
$$ \Omega = \{ \vec{x}~|~\vec{x}=(1,x_2,x_3), (x_1,1,x_3) 
~~~\mathrm{or}~~~(x_1,x_2,1) \}$$
all time-like links $U_{\vec{x},t;4}$ are frozen to constant group elements.  
For definiteness we have used embedded pure Abelian link variables
$U_{\vec{x},t;4} \equiv \cos \theta + i \sigma_3 \sin \theta.$
In the confinement phase at $\beta \le \beta_c$ we require 
$L(\vec{x}) = 0 = \langle L \rangle$ 
(corresponding to holonomy parameter $\omega=1/4$)
which is satified by $\theta = \pi / 2 N_t$. 
As in \cite{IMMPV1} we have studied also the
deconfinement case ($\beta > \beta_c$). In this case we
fixed the boundary time-like links
such that $L(\vec{x}) = \langle L \rangle $ 
for $\vec{x} \in \Omega$. In both cases 
$\langle L \rangle$ denotes 
the ensemble average $< |\sum_{\vec{x}} L(\vec{x})|/N_s^3 >$
of the volume-averaged Polyakov loop.

Each kind of boundary conditions will be employed both for the Monte Carlo
production of configurations and for their subsequent cooling.

\section{Dyonic lumps and other objects seen on the lattice}
\label{sect:simple_lumps}
In a first part of our investigation
we have searched for topologically non-trivial objects with
lowest possible action, late in the cooling history, in order to find 
systematic dependences of the selected solutions
on the spatial boundary conditions and on the temperature
of the original Monte Carlo ensemble.
Cooling was stopped at some ($n$-th) cooling iteration step when the following
criteria for the action $S_n$ were simultaneously fulfilled:
\begin{itemize}
\item $S_n < 1.5~S_{\mathrm{inst}}$,
\item $|S_n - S_{n-1}| < 0.01 S_{\mathrm{inst}}$, 
\item $S_n-2~S_{n-1}+S_{n-2} < 0$.
\end{itemize}
The last condition means that cooling just passed a point of inflection.

For each $\beta \le \beta_c$ we have scanned the topological content of
$O(200)$ configurations. In this late stage of cooling we find approximate
classical solutions which are more or less static in time 
besides non-static ones.

For both kinds of boundary conditions, among the solutions we have found 
there are such which can be easily identified as $CAL$ and
$DD$. In order to
allow a simple distinction between the
non-static calorons $CAL$ and the dissociated but static $DD$ pairs
we introduce a quantity which represents for an Euclidean configuration
the mobility of the action density.
For brevity, we call it {\it non-stationarity}
\eq \label{eq:nonstat}
\delta_t = \sum_{t,\vec{x}} \sum_{\mu < \nu}
           | s(\vec{x},t;\mu,\nu) - s(\vec{x},t-1;\mu,\nu) | / S \, .
\en
The action density per plane $s(\vec{x},t;\mu,\nu)$
and the normalization factor, the total action $S$, 
are defined in (\ref{eq:action}). 
We have monitored how frequently objects with given $\delta_t$ are found
at the lowest action plateaus.
The histograms of $\delta_t$ look similar for both types of boundary conditions.
For $\beta=2.2$ they 
have a peak at $\delta_t = 0.02-0.04$ and a long tail for large $\delta_t$.

We have convinced ourselves that a cut $\delta_t < 0.17$ well separates 
$DD$-objects which are static with two
well-separated maxima of the densities of the topological charge
$q_t({\vec x})$ and action ${\varsigma}({\vec x})$). For $\delta_t > 0.17$ the
objects can be easily interpreted as $CAL$ which are non-static, with an
approximately $O(4)$ rotationally symmetric action distribution,
with a single maximum of $q_t({\vec x})$ and ${\varsigma}({\vec x})$ )
in $3D$ space. Both $DD$ and $CAL$ are
showing a pair of opposite-sign peaks of the Polyakov loop.

\subsection{$DD$ pairs}

\begin{figure}[!htb]
(a)\hspace{0.5cm}\includegraphics[width=0.4\textwidth]{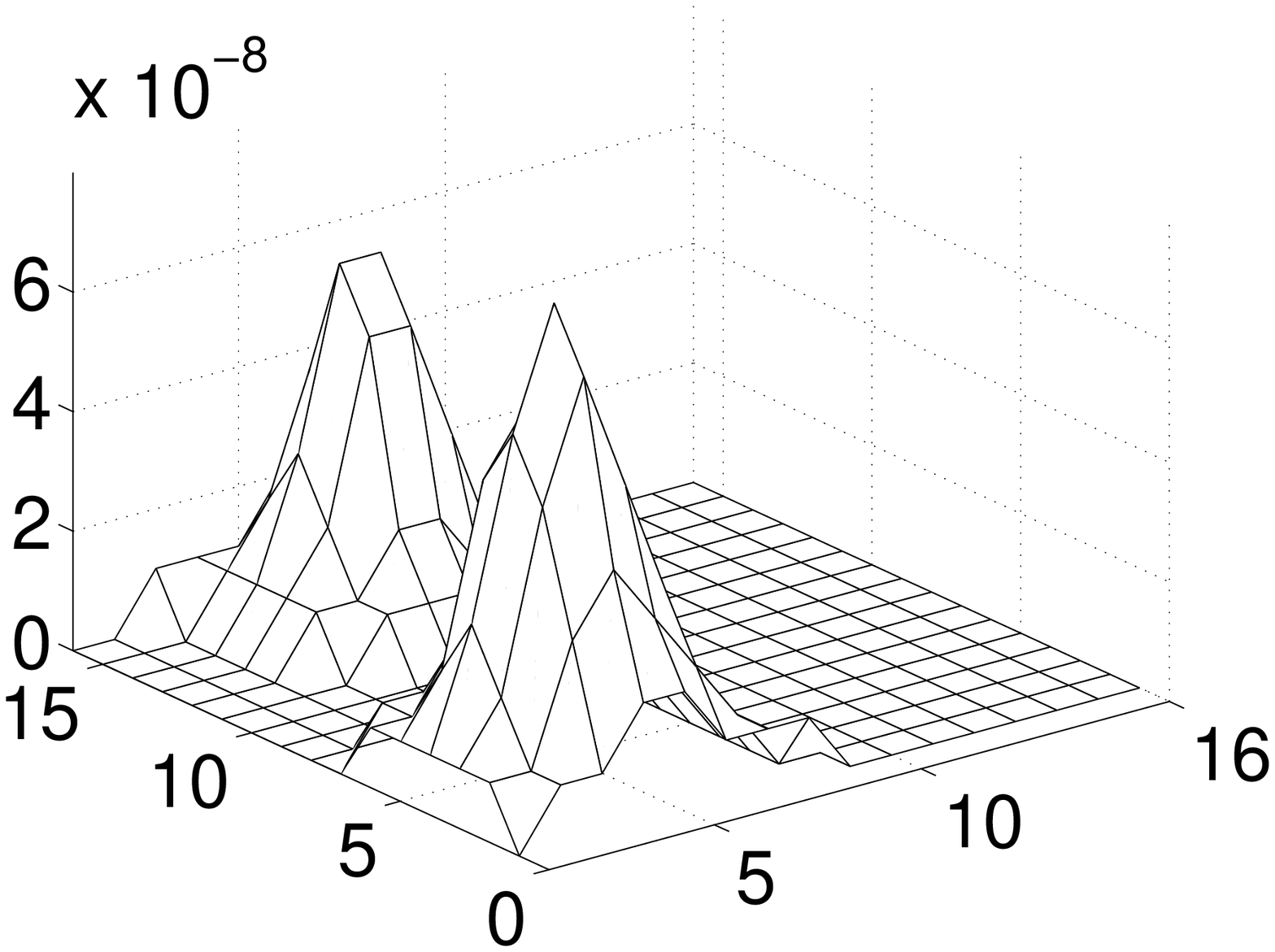}%
\hspace{0.3cm}
(b)\hspace{0.5cm}\includegraphics[width=0.4\textwidth]{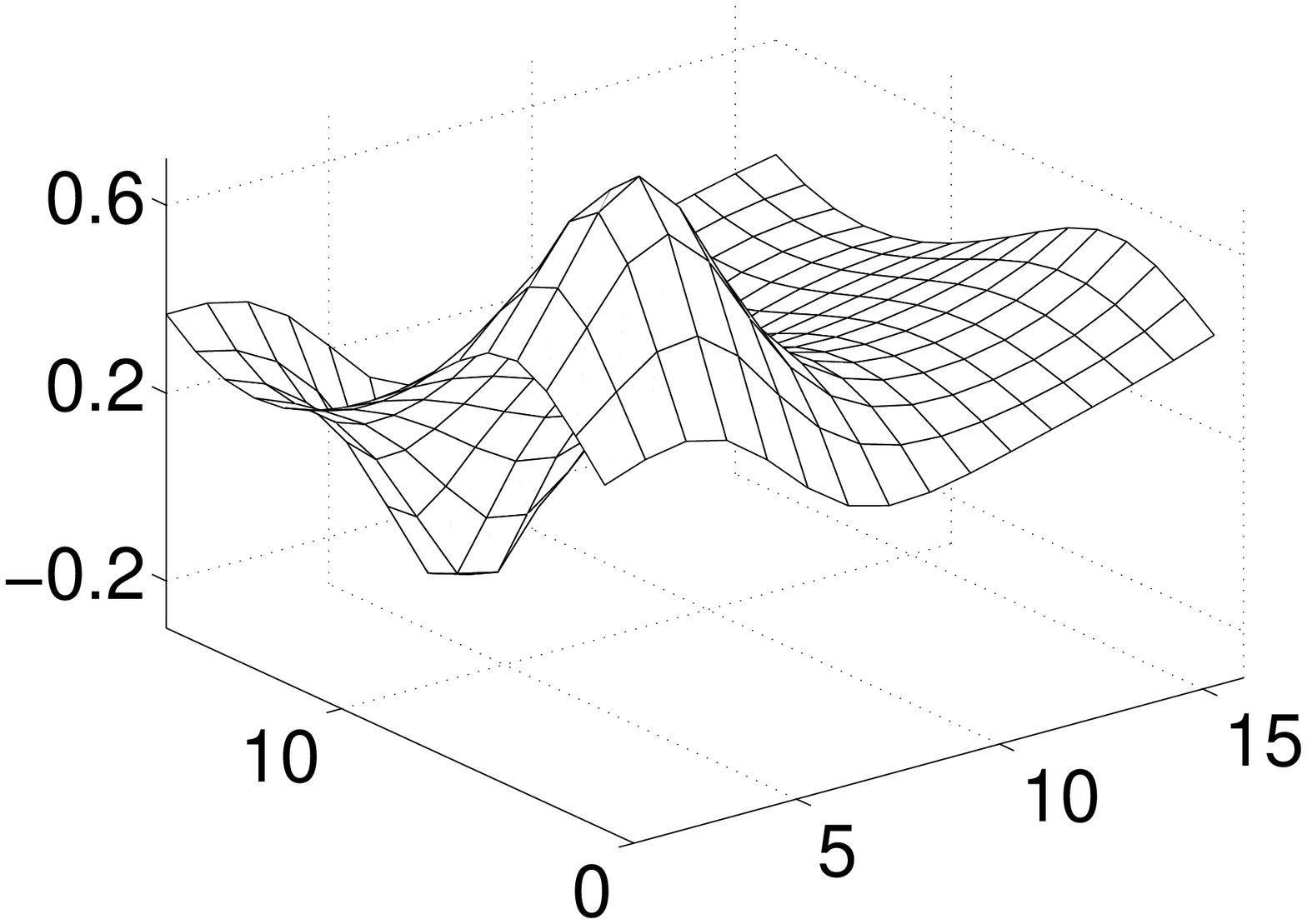}

\vspace{1cm}
(c)\hspace{0.5cm}\includegraphics[width=0.4\textwidth]{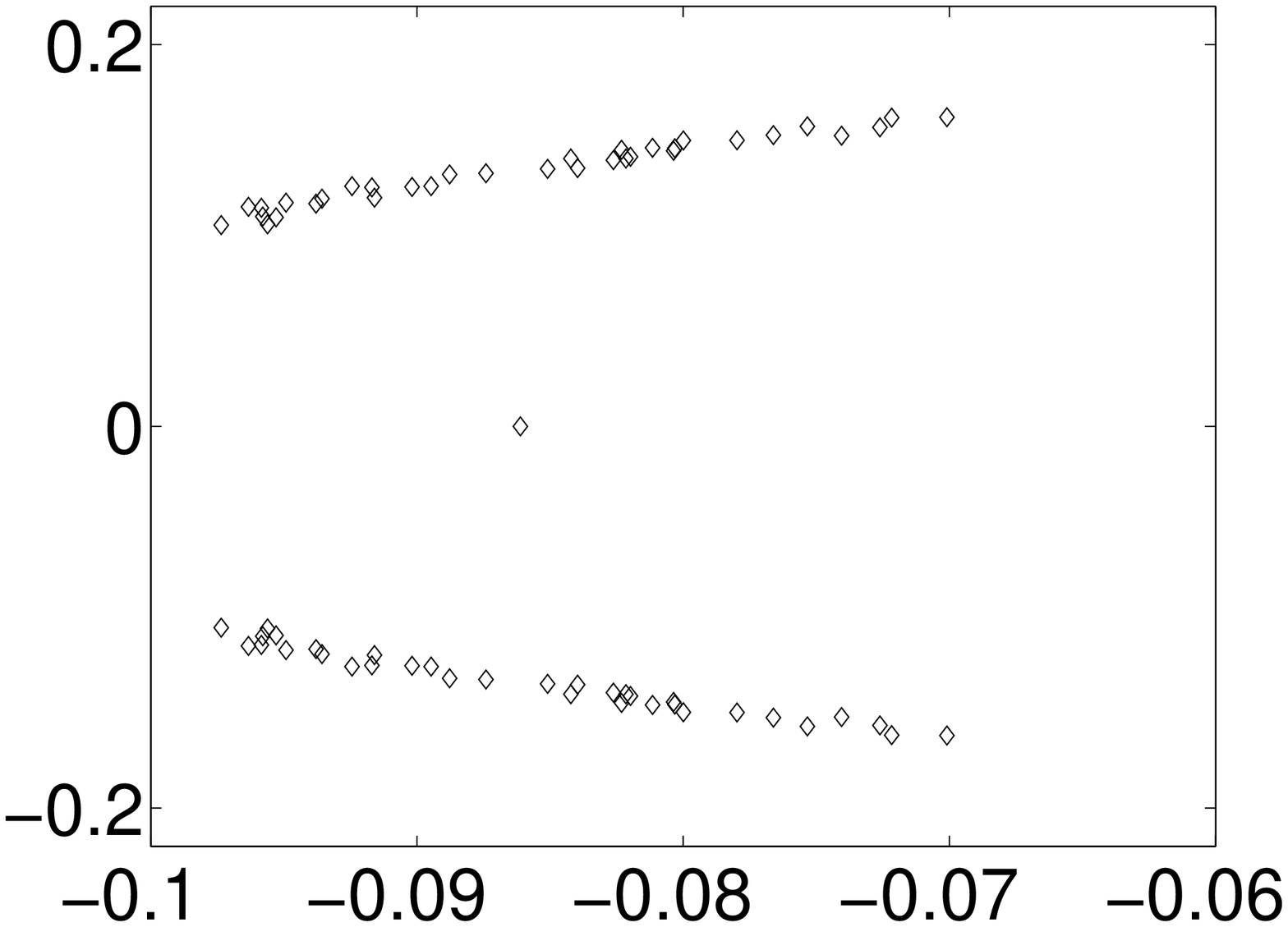}%
\hspace{0.3cm}
(d)\hspace{0.5cm}\includegraphics[width=0.4\textwidth]{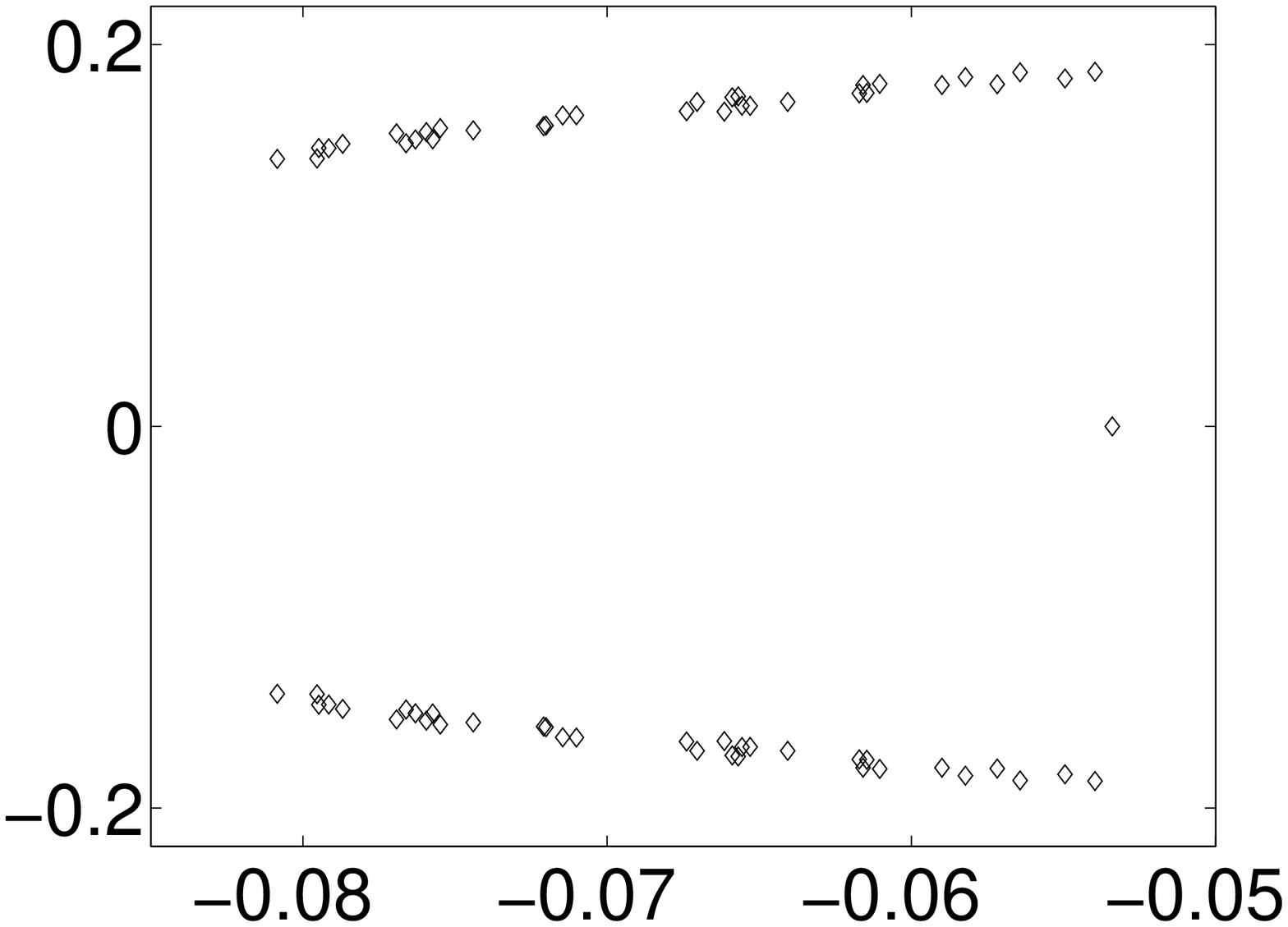}

\vspace{1cm}
(e)\hspace{0.5cm}\includegraphics[width=0.4\textwidth]{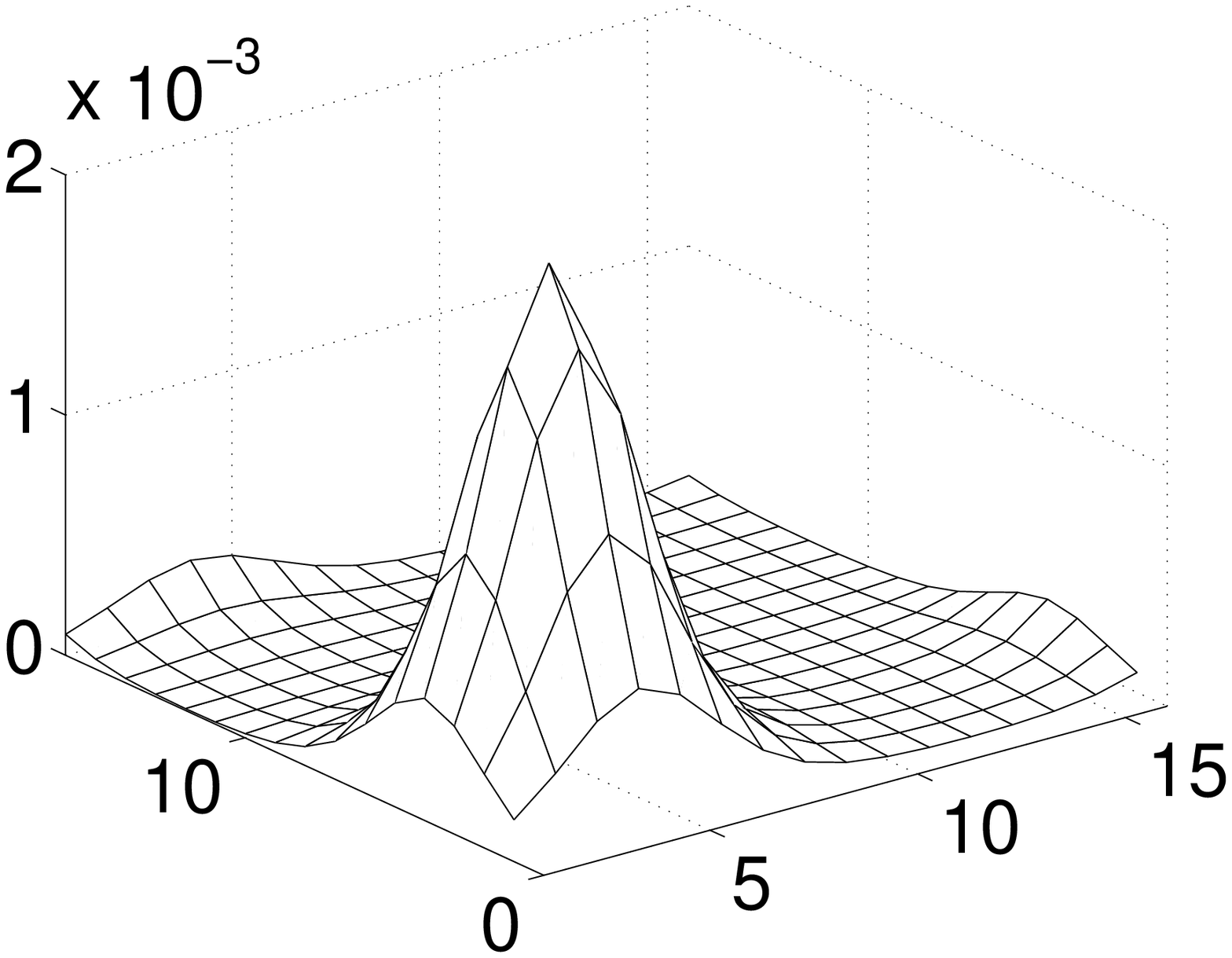}%
\hspace{0.3cm}
(f)\hspace{0.5cm}\includegraphics[width=0.4\textwidth]{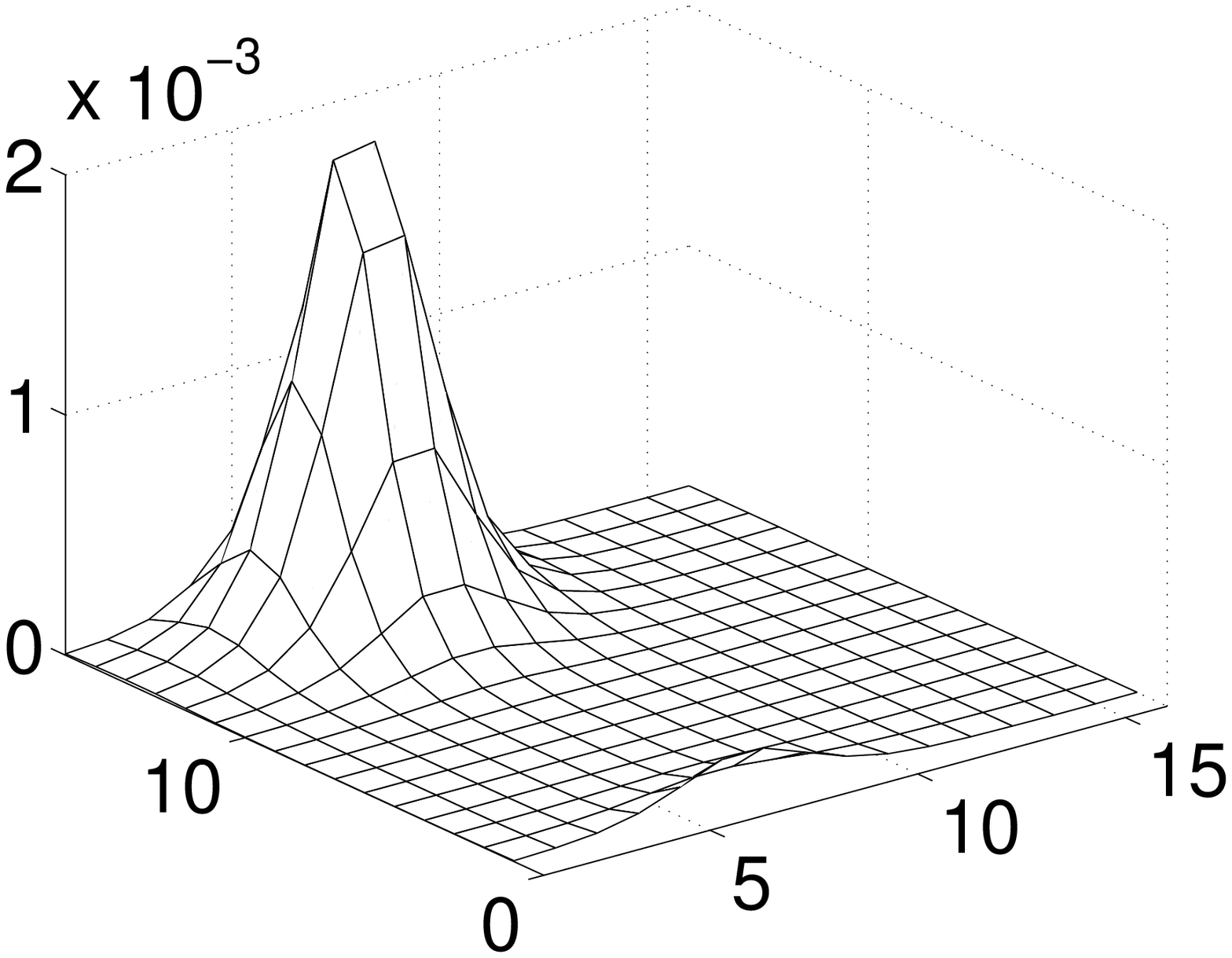}
\caption{
 Various portraits of a selfdual $DD$ pair obtained by cooling under 
 periodic gluonic boundary conditions. The sub-panels show:
 appropriate $2D$ cuts of the topological charge density (a) and of 
 the Polyakov loop (b), the plot of lowest fermionic eigenvalues (c,d) 
 and the $2D$ cut of the real-mode fermion densities (e,f), for 
 the cases of time-periodic (c,e) and time-antiperiodic (d,f) fermionic
 boundary conditions, respectively ($\beta=2.2$ and lattice size 
 $16^3\times4$).}
\label{fig:dd}
\end{figure}

For a special $DD$ solution found with p.b.c., we show in Figs. 
\ref{fig:dd} (a) and (b) two-dimensional cuts
of the topological charge density $q_t(\vec{x})$ (a) and
of the Polyakov loop distribution $L(\vec{x})$ (b). 
The $DD$ solution was obtained from an equilibrium configuration
representing $\beta=2.2$, {\it i. e.} the confinement phase. 
We clearly see the opposite-sign peaks of the Polyakov loop variable 
correlated with the equal-sign maxima of the topological charge density.
The boundary values of the Polyakov loop are slightly varying because
they are not fixed here to a well-defined value. This is the only difference 
observed between the two types of boundary conditions. In principle, 
for p.b.c. the holonomy could be arbitrary. What really happens to the 
asymptotic holonomy we will discuss in Section \ref{sect:gases}.
As a consequence the ratio of the action 
carried by the well-separated dyon constituents can take any value.

For the same $DD$ solution,
Figs. \ref{fig:dd} (c) and (d) show the scatter plot of the 70 lowest 
complex Wilson fermion eigenvalues (\ref{eq:fermionoperator}) 
for $\kappa=0.14$, both for time-periodic (c) and time-antiperiodic (d)
boundary conditions for the fermion fields. 
In both cases we find one isolated low-lying real eigenvalue
which can be related to a zero-mode of the zero-mass continuum Dirac operator.
The corresponding (projected) eigenmode densities 
$\psi^{\dagger} \psi(x)$ are drawn below, in Figs. \ref{fig:dd} 
(e) and (f).  They show a localization behavior as analytically proposed 
in Eqs. (\ref{eq:fermiondensity}). 
For the time-antiperiodic b.c. the eigenmode is localized at the dyon exhibiting
the negative peak of the Polyakov loop related to Taubes winding \cite{CKvB}.

\begin{figure}[!htb]
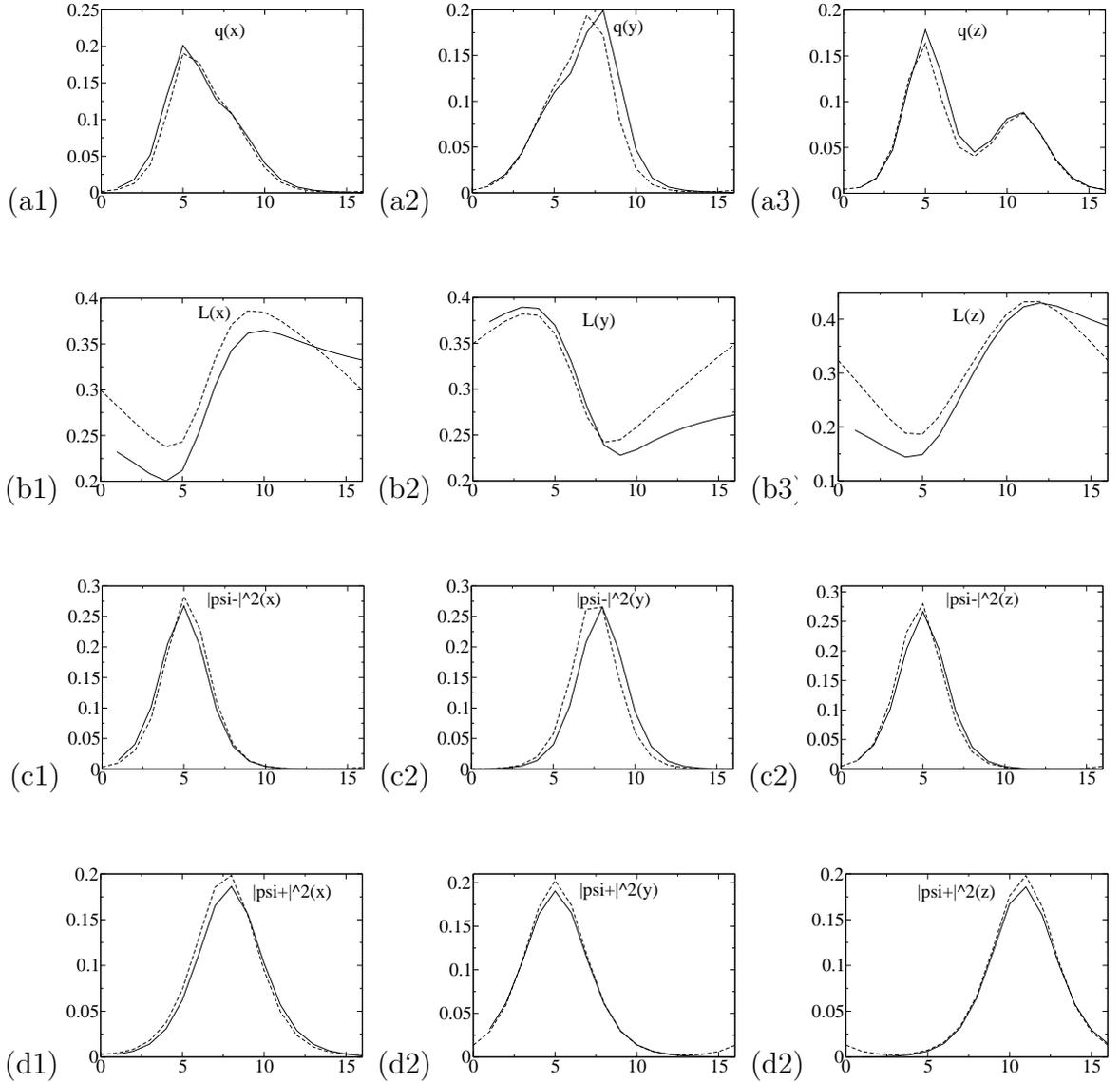

\begin{center}

(a1)\hspace{.1cm}\includegraphics[width=.26\textwidth]{ddqx.eps}%
\hspace{.2cm}(a2)\hspace{.1cm}\includegraphics[width=.26\textwidth]{ddsy.eps}%
\hspace{.2cm}(a3)\hspace{.1cm}\includegraphics[width=.26\textwidth]{ddqz.eps}

\vspace{1cm}

(b1)\hspace{.1cm}\includegraphics[width=.26\textwidth]{ddlx.eps}%
\hspace{.2cm}(b2)\hspace{.1cm}\includegraphics[width=.26\textwidth]{ddly.eps}%
\hspace{.2cm}(b3)\hspace{.1cm}\includegraphics[width=.26\textwidth]{ddlz.eps}

\vspace{1cm}

(c1)\hspace{.1cm}\includegraphics[width=.26\textwidth]{ddfapbx.eps}%
\hspace{.2cm}(c2)\hspace{.1cm}\includegraphics[width=.26\textwidth]{ddfapby.eps}%
\hspace{.2cm}(c2)\hspace{.1cm}\includegraphics[width=.26\textwidth]{ddfapbz.eps}

\vspace{1cm}

(d1)\hspace{.1cm}\includegraphics[width=.26\textwidth]{ddfpbx.eps}%
\hspace{.2cm}(d2)\hspace{.1cm}\includegraphics[width=.26\textwidth]{ddfpby.eps}%
\hspace{.2cm}(d2)\hspace{.1cm}\includegraphics[width=.26\textwidth]{ddfpbz.eps}

\vspace{1cm}
\end{center}
\caption{The $DD$ lattice configuration of Fig. 1 (dashed lines), fitted 
 by the KvB solution (solid lines) according to the action density, 
 shown in three spatial views ($1D$ projections). 
 In (a1),(a2),(a3) the topological charge density is summed over the two 
 unspecified spatial coordinates.
 In (b1),(b2),(b3) the Polyakov loop is averaged over the two unspecified 
 spatial coordinates. 
 In the time-antiperiodic case (c1,c2,c3) or the time-periodic case 
 (d1,d2,d3), respectively, the fermion density it is summed over the two 
 unspecified spatial coordinates.}
\label{fig:dd_fit}
\end{figure}

For the given solution created on the lattice we have carried out a fit with 
the analytic formula \cite{KvB} to reproduce
the action density (\ref{eq:solutionact}). 
This has provided the parameter values
$\vec{x}_1=(8, 5, 11 )$, $\vec{x}_2=(5, 8, 5 )$ and $\omega = 0.202$.
Fig. \ref{fig:dd_fit} shows one-dimensional projections of the same gauge field
configuration together with the analytical results produced with the given fit 
parameters for the topological density,
the Polyakov loop and the fermionic mode density with time-antiperiodic
and -periodic boundary conditions according to the expressions 
(\ref{eq:fermiondensity}). The last two curves are predictions, rather
than fits. There is an impressive agreement with the numerical shape
of the fermionic zero-mode density.

Gauge fixing to MAG we can search for the 
Abelian monopole content of the field configurations under inspection.
We are interested in the positions of the world lines of
monopole-antimonopole pairs. For static $DD$ solutions
we always find a pair of static (anti)monopoles with world lines
coinciding with the centers of the dyons. 
All these features 
have been observed for $DD$-solutions irrespective of the spatial boundary 
conditions imposed in the process of cooling.

\subsection{$CAL$ configurations}

\begin{figure}[!htb]
(a)\hspace{0.5cm}\includegraphics[width=0.4\textwidth]{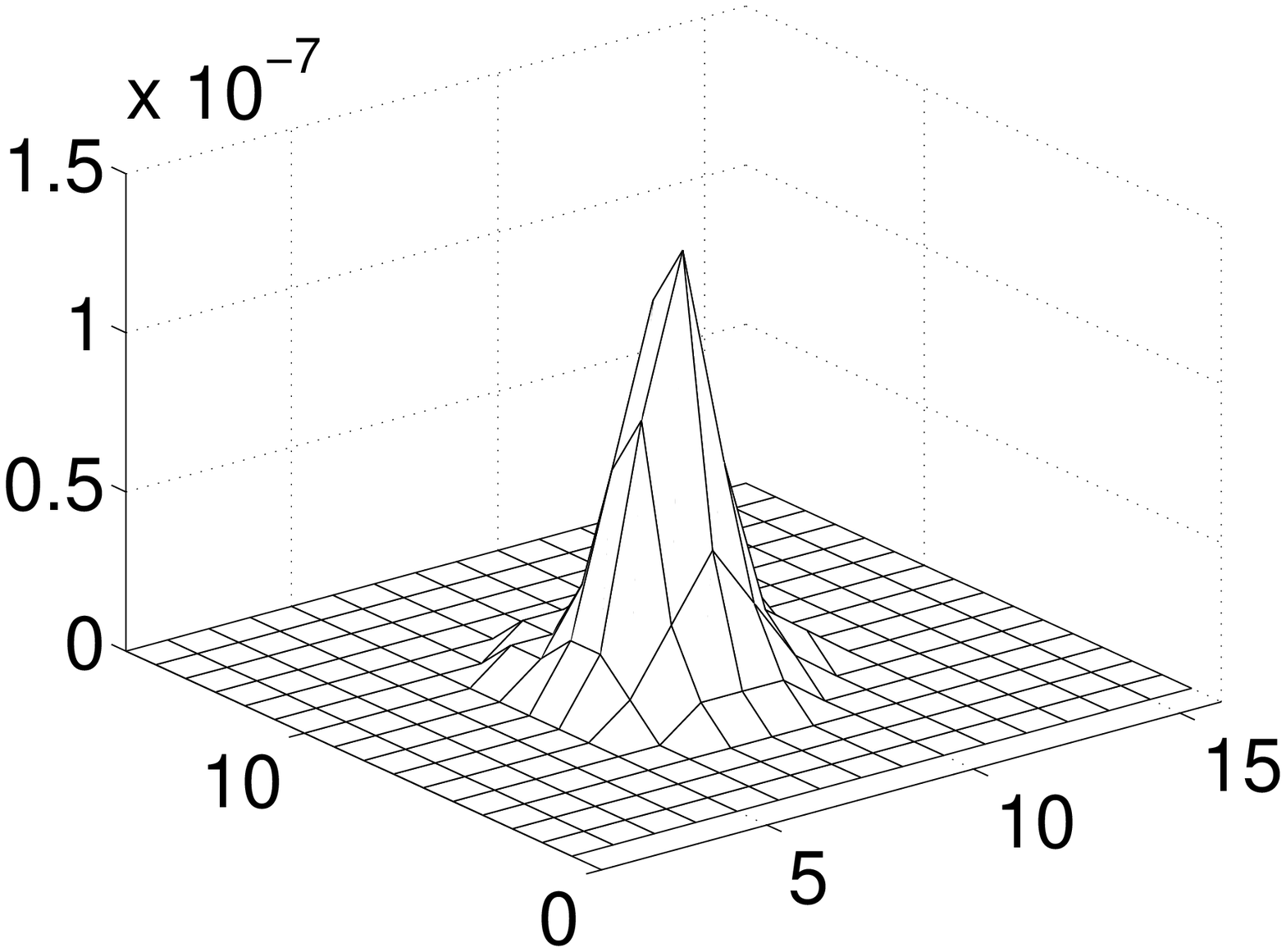}%
\hspace{0.3cm}
(b)\hspace{0.5cm}\includegraphics[width=0.4\textwidth]{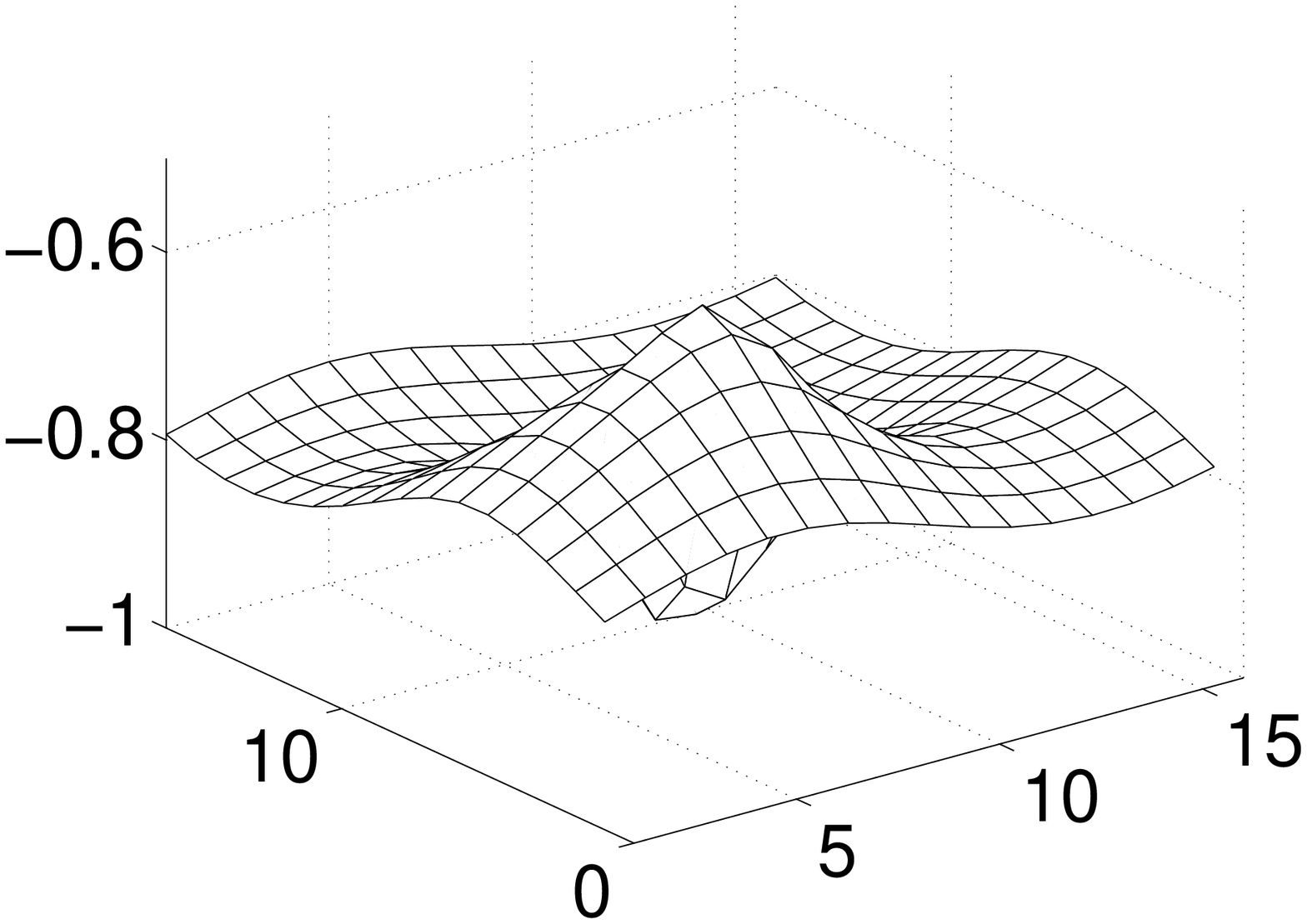}

\vspace{1cm}
(c)\hspace{0.5cm}\includegraphics[width=0.4\textwidth]{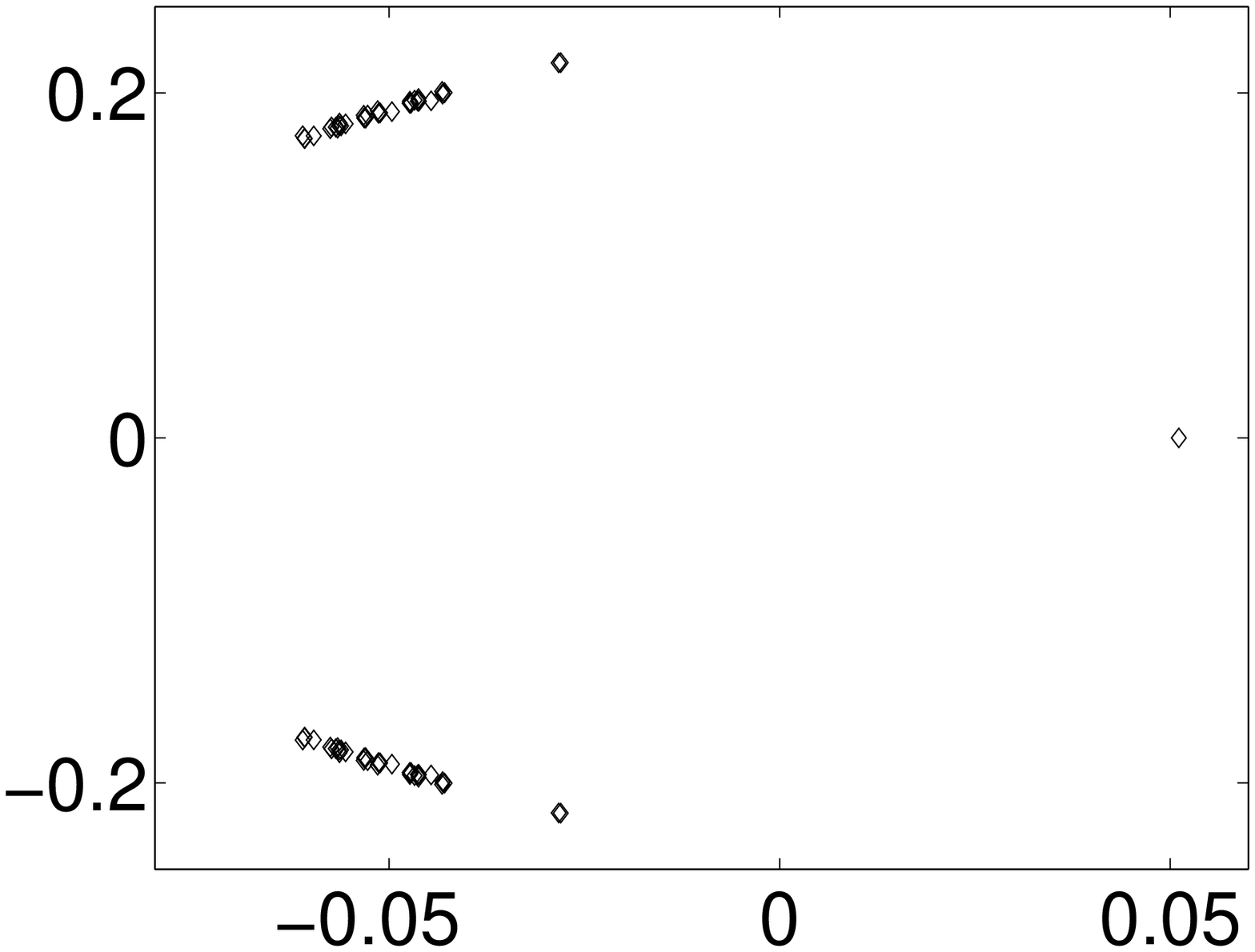}%
\hspace{0.3cm}
(d)\hspace{0.5cm}\includegraphics[width=0.4\textwidth]{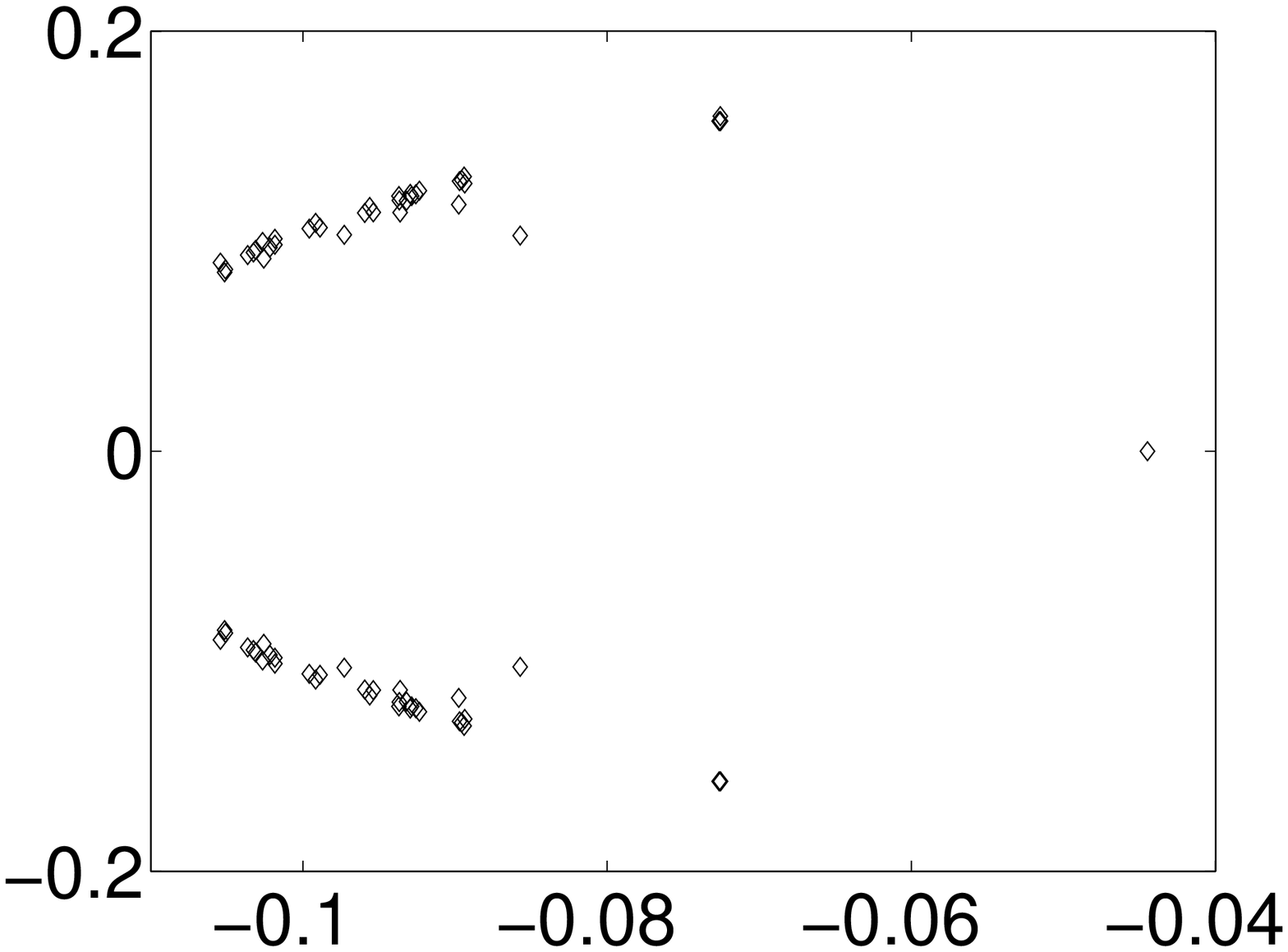}

\vspace{1cm}
(e)\hspace{0.5cm}\includegraphics[width=0.4\textwidth]{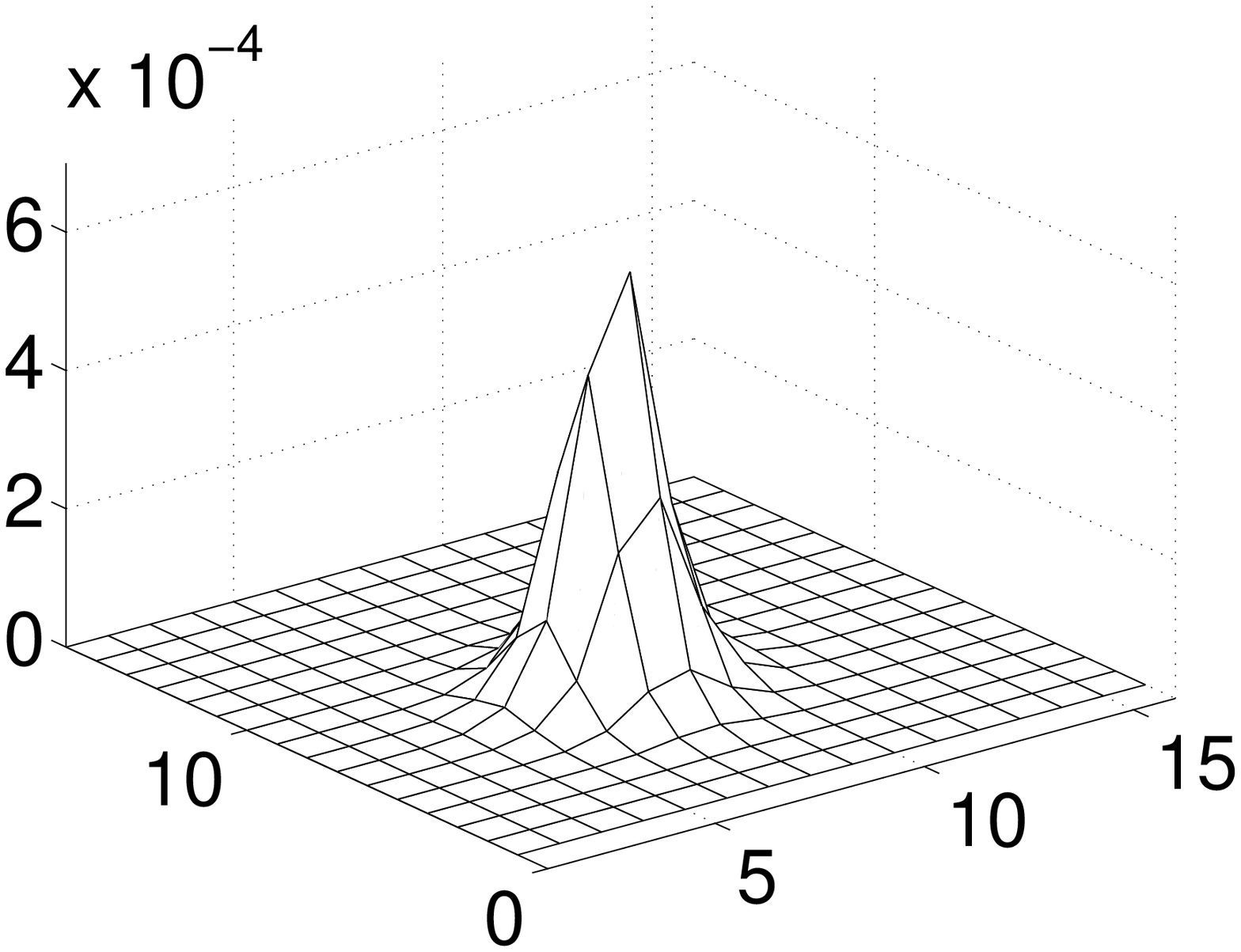}%
\hspace{0.3cm}
(f)\hspace{0.5cm}\includegraphics[width=0.4\textwidth]{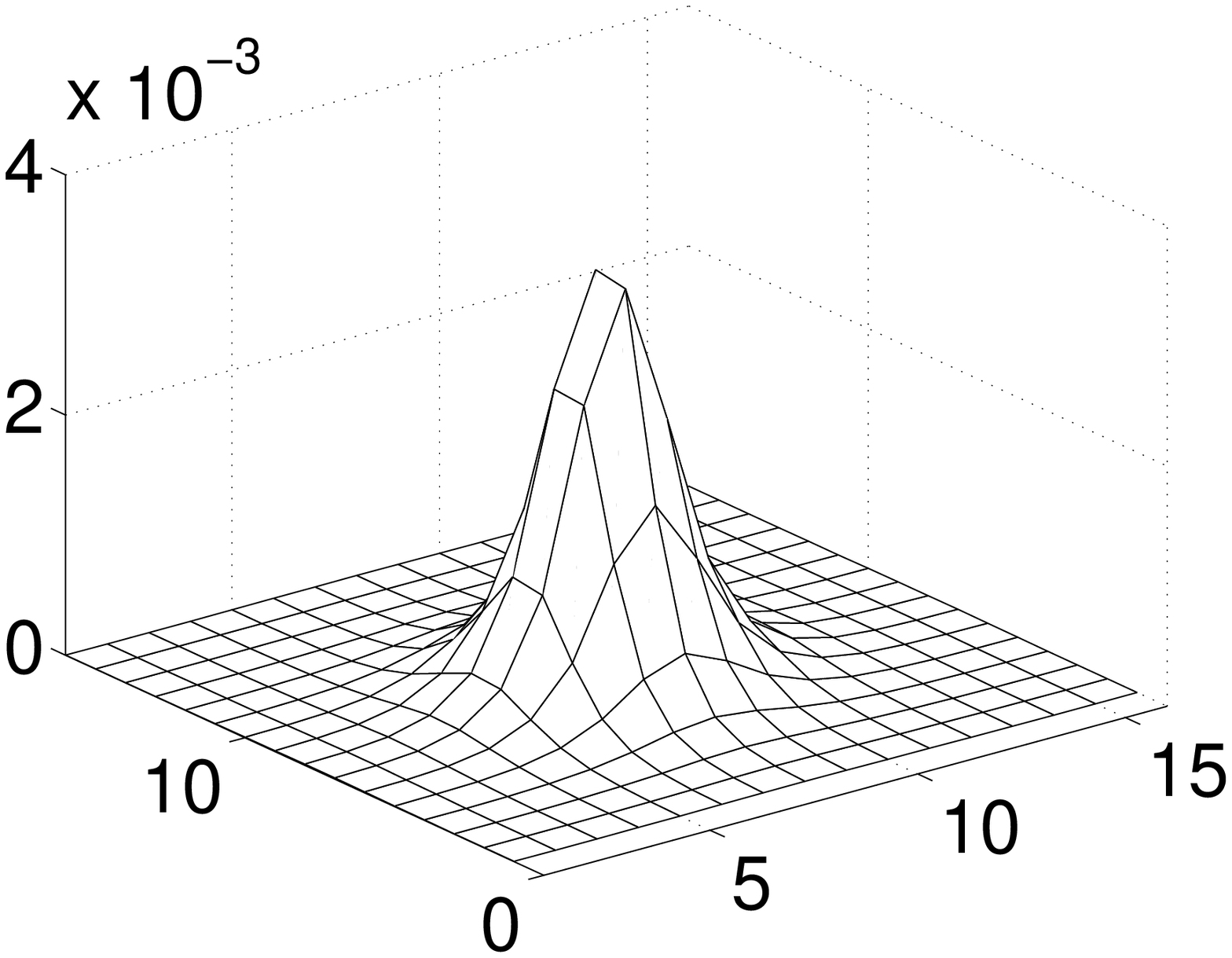}
\caption{
 Various portraits of a selfdual $CAL$ configuration obtained by cooling 
 under periodic gluonic boundary conditions. The sub-panels show:
 appropriate $2D$ cuts of the topological charge density (a) and of 
 the Polyakov loop (b), the plot of lowest fermionic eigenvalues (c,d)
 and the $2D$ cut of the real-mode fermion densities (e,f), for 
 the cases of time-periodic (c,e) and time-antiperiodic (d,f) fermionic
 boundary conditions, respectively ($\beta=2.2$ and lattice size 
 $16^3\times4$).}
\label{fig:cal}
\end{figure}

In Fig. \ref{fig:cal} we show a typical $CAL$ solution, with an approximately 
$4D$ rotationally invariant action distribution, obtained 
at $\beta=2.2$ from cooling with periodic boundary conditions.
The configuration possesses a large value of the non-stationarity
$\delta_t$. Again we plot 
$2D$ cuts for the topological charge density,
for the Polyakov loop and the fermionic eigenvalues
together with the eigenmode density for the distinct real eigenvalue.
The full topological charge $Q_t$ is unity. 
The expected pair of narrow opposite-sign peaks of the Polyakov loop 
is nicely visible.

The fermionic zero-modes for time-periodic and time-antiperiodic b.c.
for this configuration are only slightly shifted relative to each other.
A reasonable fit with the
analytic solution can be obtained showing that this caloron is nothing
but a limiting case of the generic $DD$ solutions. 
At that point we may conclude that cooling, even with non-fixed holonomy,
yields almost-classical solutions which show all characteristica of the 
KvB calorons. The typical $CAL$ configurations show,
after putting them into MAG, a closed Abelian monopole loop 
circulating around the maximum of the action density in the $4D$-space.

\subsection{$D \bar{D}$ pairs}

As previously observed for the case with f.h.b.c. \cite{IMMPV1},
we have found also other field configurations with an action on the 
instanton level, $S \simeq S_{\mathrm{inst}}$, which are
very stable against cooling which motivates us
to interprete them also as approximate solutions
of the lattice equations of motion. With very low non-stationarity
$\delta_t = 0.004 \pm 0.002$, they consist of two lumps
of action with opposite-sign topological charge densities. We call
them dyon-antidyon pairs, $D \bar{D}$.
Each of their lumps turns out approximately (anti)self-dual. The total
topological charge of the entire configuration is always zero. Therefore, 
each of the lumps carries
half-integer topological charge. The Polyakov loop exhibits two peaks,
in this case of equal sign.

Also for $D \bar{D}$ pairs MAG fixing offers an Abelian monopole interpretation.
After Abelian projection a static Abelian monopole-antimonopole
pair can be found at the positions of the topological charge centers.

\begin{figure}[!htb]
(a)\hspace{0.5cm}\includegraphics[width=0.4\textwidth]{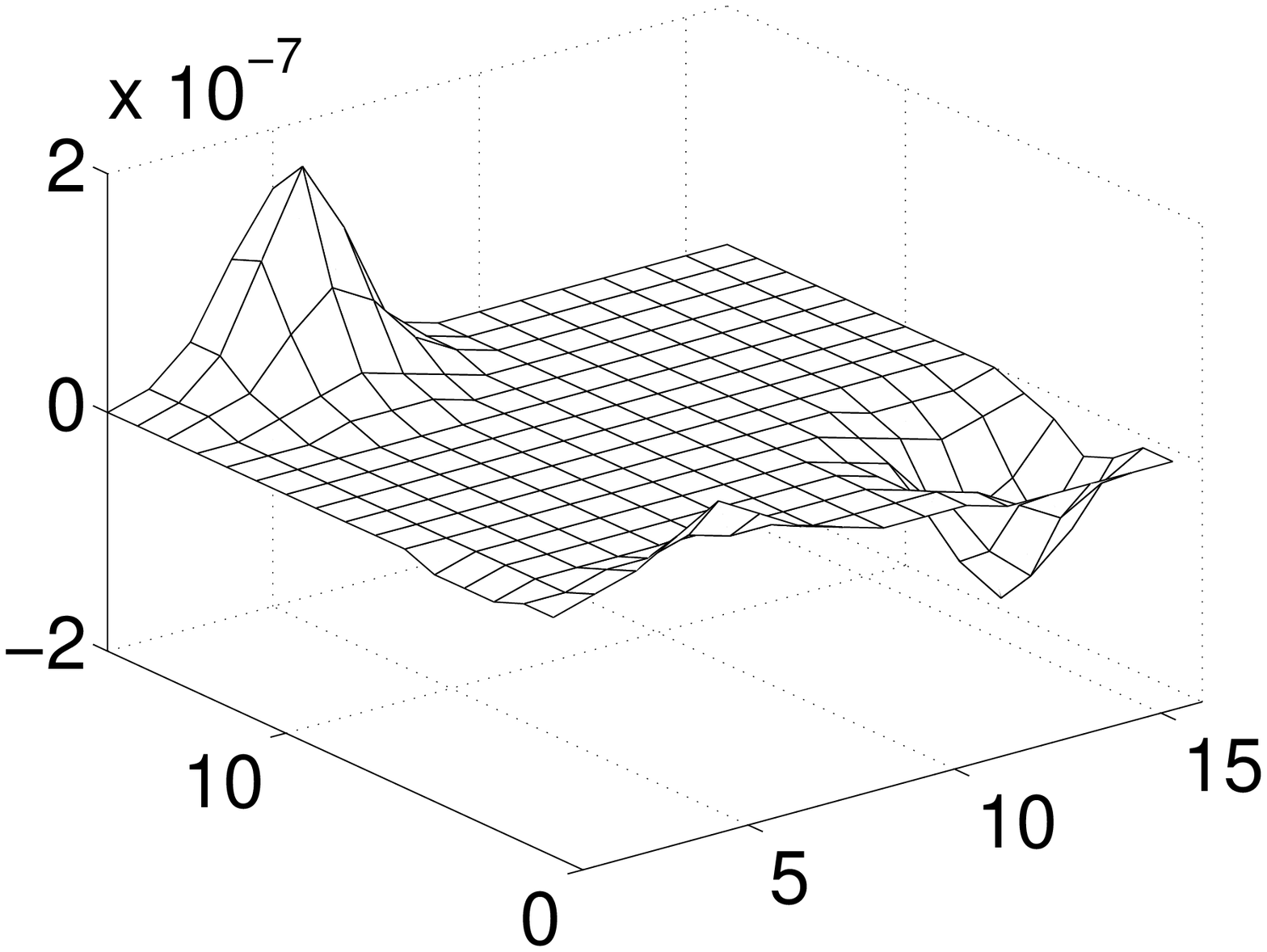}%
\hspace{0.3cm}
(b)\hspace{0.5cm}\includegraphics[width=0.4\textwidth]{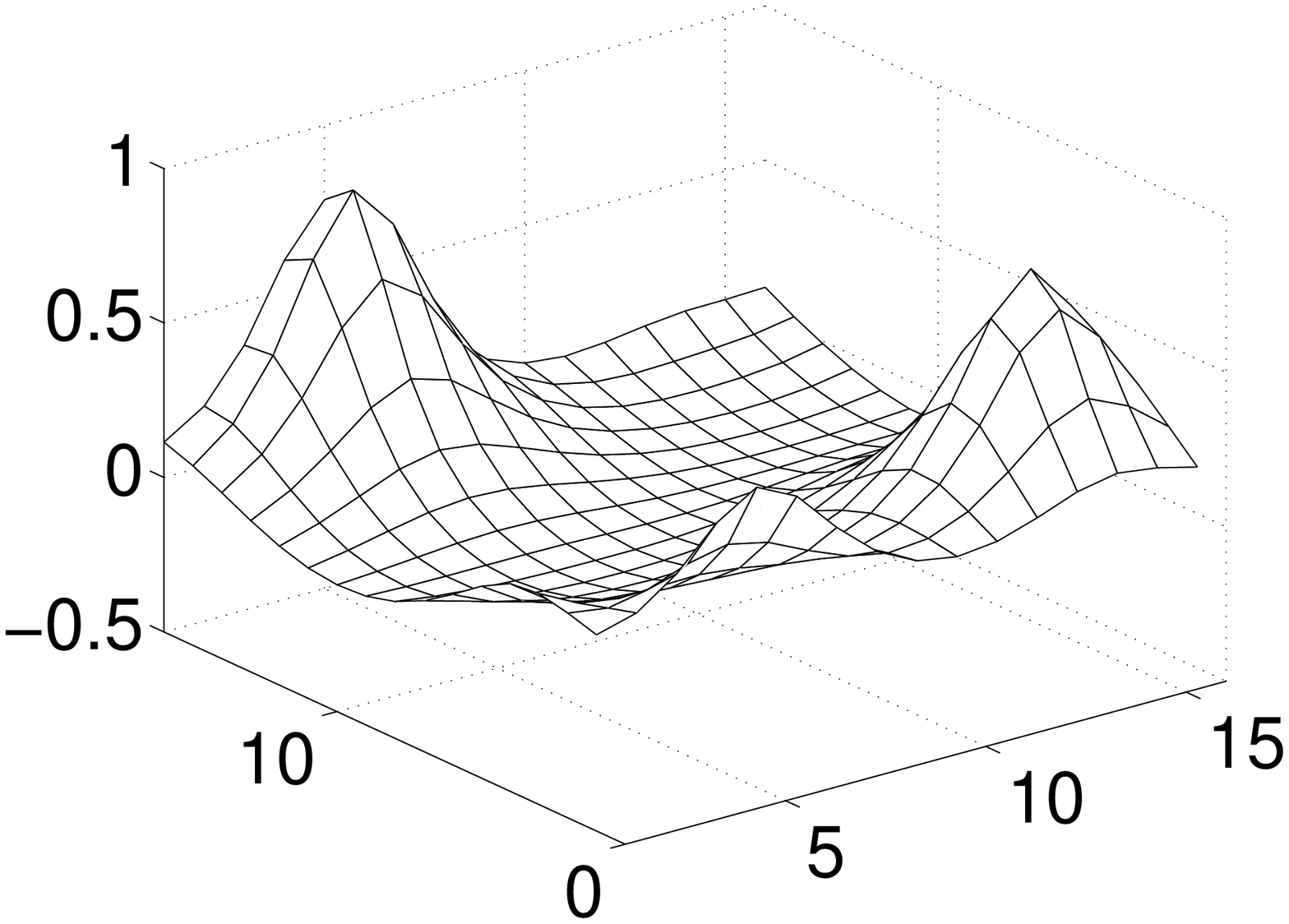}

\vspace{1cm}
(c)\hspace{0.5cm}\includegraphics[width=0.4\textwidth]{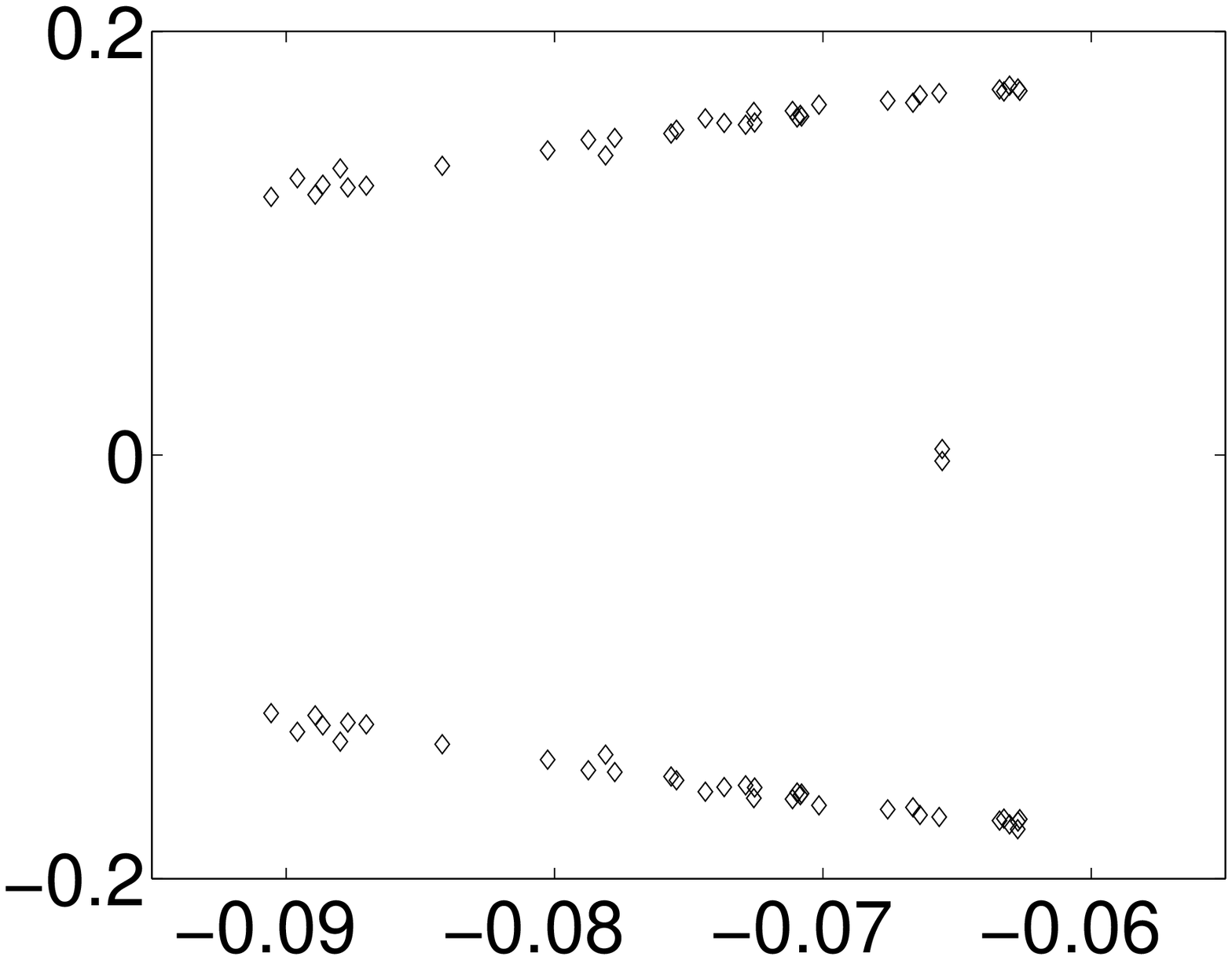}%
\hspace{0.3cm}
(d)\hspace{0.5cm}\includegraphics[width=0.4\textwidth]{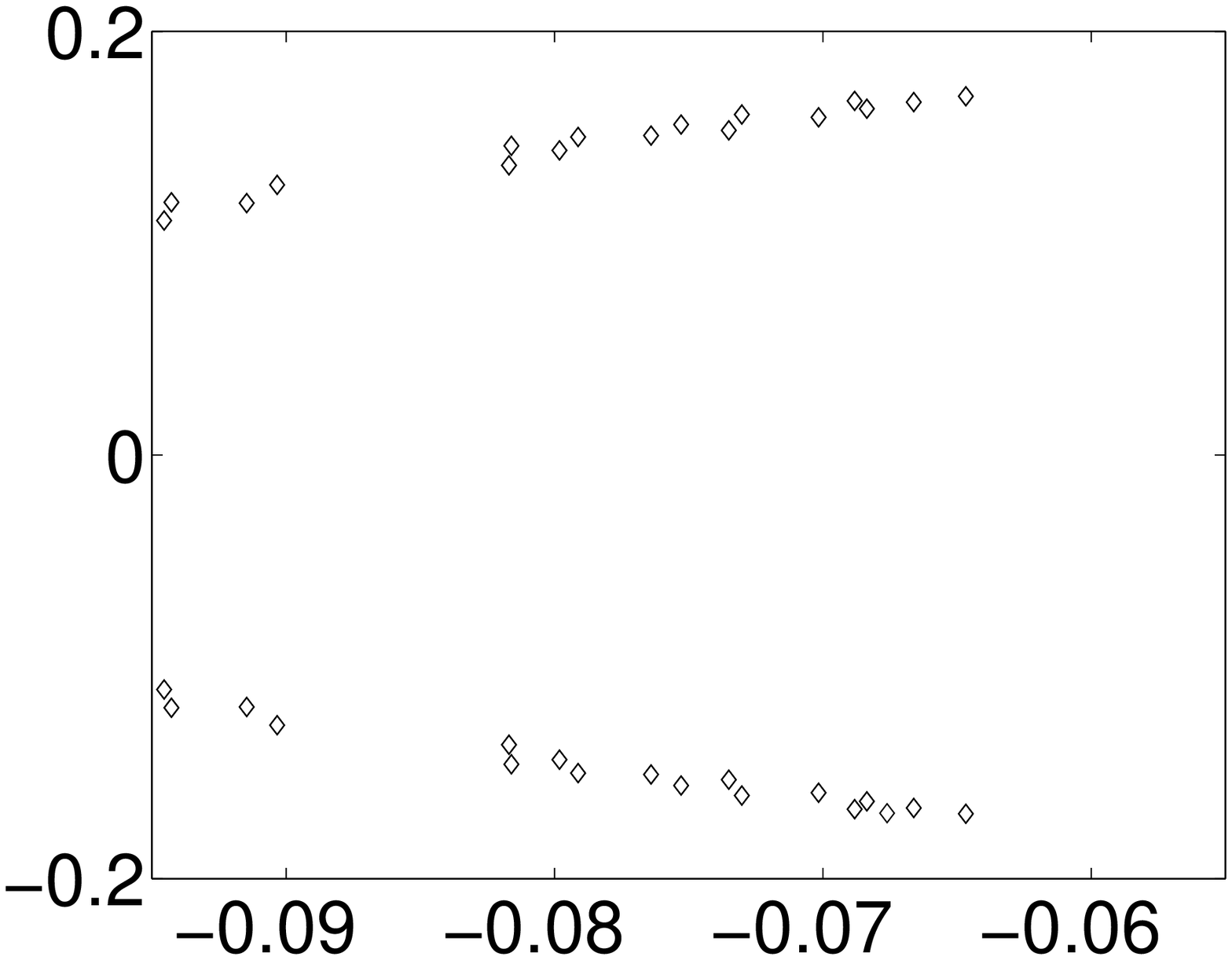}

\vspace{1cm}
(e)\hspace{0.5cm}\includegraphics[width=0.4\textwidth]{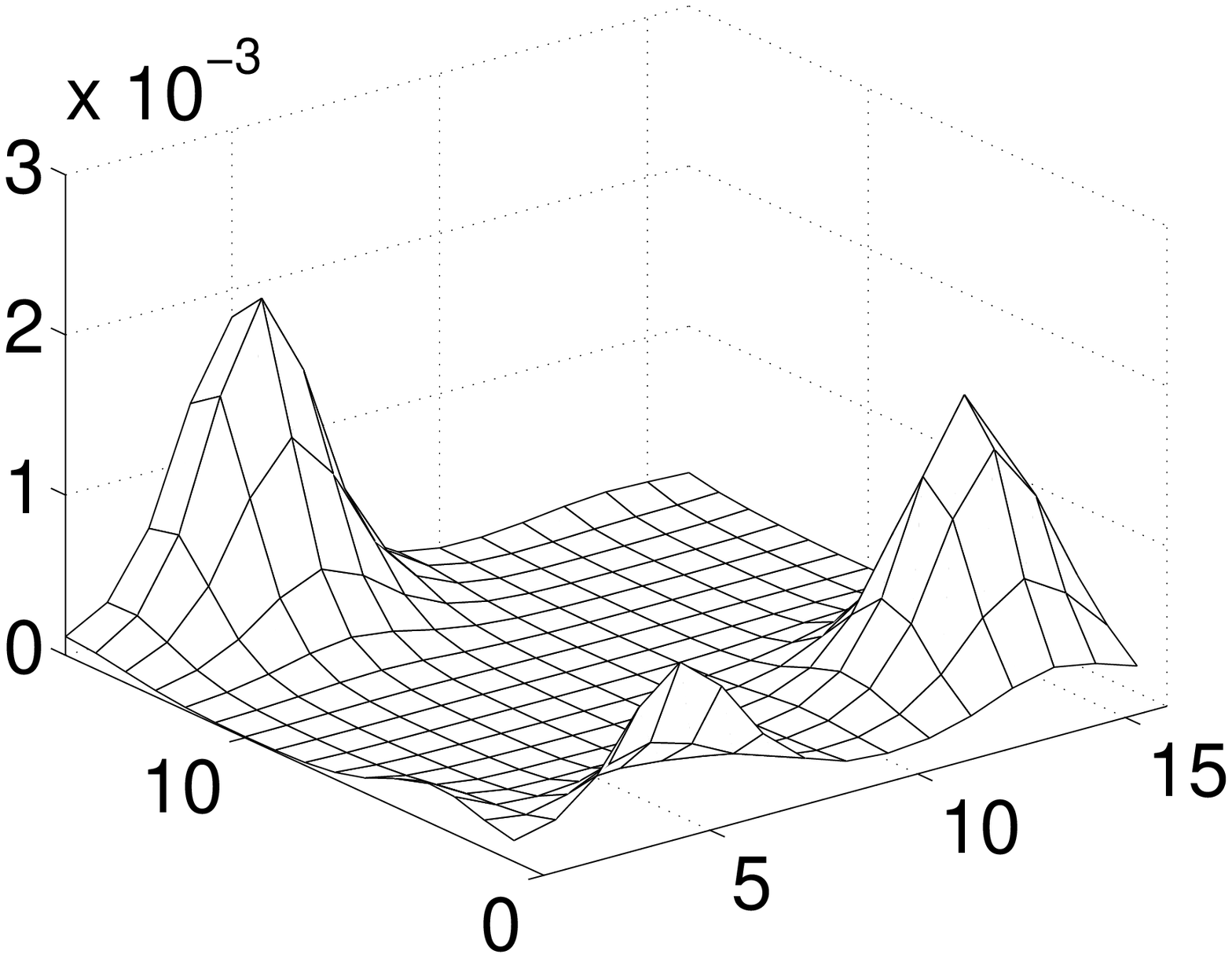}
\caption{
 Various portraits of a mixed-duality $D \bar{D}$ pair obtained by cooling 
 under periodic gluonic boundary conditions. The sub-panels show:
 appropriate $2D$ cuts of the topological charge density (a) and of 
 the Polyakov loop (b), the plot of lowest fermionic eigenvalues (c,d) for
 the cases of time-periodic (c) and time-antiperiodic (d) fermionic
 boundary conditions, respectively ($\beta=2.2$ and lattice size 
 $16^3\times4$). A $2D$ cut of the fermionic mode density related to
 the two distinct {\it almost} real eigenvalues in (c) is shown in (e).}
\label{fig:dad}
\end{figure}

Searching for the eigenvalues of the Wilson-Dirac operator we did not
find real eigenvalues but sometimes pairs of complex conjugated
eigenvalues with very small 
imaginary parts. This feature is very similar to that
of dilute superpositions of instantons with antiinstantons. This gives us good
reason to interpret $D \bar{D}$ pairs as superpositions of single BPS
solutions with half-integer topological charge. To the best of 
our knowledge, analytic solutions of this kind have not been reported 
in the literature.
An example for a $D \bar{D}$ pair is reproduced in Fig. \ref{fig:dad}.
We did not find any real or near-to-real modes for the time-antiperiodic 
boundary conditions.

\subsection{The composition of the lowest action plateaus}

Finally, by cooling with both kinds of spatial boundary conditions
we have found objects becoming very stable at even lower action,
for lattice size $16^3 \times 4$,
$S \simeq S_{\mathrm{inst}}/2$ and $\simeq S_{\mathrm{inst}}/4$.
Their (color-) electric contribution to the action is very small compared
with the magnetic contribution.
Moreover, they are perfectly static with $\delta_t = 0.003 \pm 0.002$.
Employing MAG we have convinced ourselves that they are purely Abelian.
In the confinement phase they are occuring quite rarely directly in the 
cooling process. They are more common to appear after monopole-antimonopole 
pairs observed at $S \approx S_{\mathrm{inst}}$ annihilate in the final stage 
of the relaxation. Therefore, we shall
not consider them in detail here. But they seem to play an important r\^ole
in the deconfinement phase \cite{LS,IMMPV2}. Since they are purely magnetic
solutions - pure magnetic fluxes or 't Hooft-Polyakov-like monopoles -
we shall abbreviate them $M$.

In Table 1 the relative frequencies to find different types
of classical configurations ($DD$, $CAL$, $D \bar{D}$ and $M$)
at and below the one-instanton action plateaus are shown. 
We compare here p.b.c. with f.h.b.c. For each $\beta \le \beta_c$ we have 
investigated $O(200)$ Monte Carlo equilibrium configurations.
\begin{table}[ht]
\caption{Relative frequencies of the occurrence of different kinds of
(approximate) solutions, for different values of $\beta$ and for both 
kinds of boundary conditions of the gauge field 
(f.h.b.c. and standard p.b.c.). The lattice size is $16^3 \times 4$.}
\vspace{5mm}
\begin{center}
\begin{tabular}{lcccc}
\hline
Type of & Boundary  & $\beta=2.20 $   & $ \beta=2.25 $   & $\beta=2.30
                                                        \simeq \beta_c    $  \\
solution        & condition &                 &                  &                    \\
\hline
$DD$            & f.h.b.c.  & $0.46 \pm 0.05 $ & $ 0.52 \pm 0.05$ & $ 0.45 \pm 0.05 $  \\
                & p.b.c.    & $0.43 \pm 0.05 $ & $ 0.44 \pm 0.05$ & $ 0.23 \pm 0.03 $  \\
\hline
$CAL$           & f.h.b.c.  & $0.19 \pm 0.03 $ & $ 0.17 \pm 0.03$ & $ 0.15 \pm 0.03 $  \\
                & p.b.c.    & $0.24 \pm 0.03 $ & $ 0.26 \pm 0.03$ & $ 0.26 \pm 0.03 $  \\
\hline
$D\overline{D}$ & f.p.b.c.  & $0.28 \pm 0.04 $ & $ 0.26 \pm 0.04$ & $ 0.31 \pm 0.04 $  \\
                & p.b.c.    & $0.18 \pm 0.03 $ & $ 0.16 \pm 0.03$ & $ 0.10 \pm 0.02 $  \\
\hline
$M$             & f.p.b.c.  & $0.01 \pm 0.01 $ & $ 0.01 \pm 0.01$ & $ 0.03 \pm 0.01 $  \\
                & p.b.c.    & $0.04 \pm 0.02 $ & $ 0.03 \pm 0.01$ & $ 0.10 \pm 0.02 $  \\
\hline
trivial vacuum  & f.p.b.c.  & $0.06 \pm 0.02 $ & $ 0.04 \pm 0.02$ & $ 0.06 \pm 0.02 $  \\
                & p.b.c.    & $0.11 \pm 0.02 $ & $ 0.11 \pm 0.02$ & $ 0.31 \pm 0.04 $  \\
\hline
\end{tabular}
\end{center}
\end{table}
\noi
We can conclude that cooling applied to configurations in the confinement phase 
produces all objects with relative probabilities which are approximately 
independent of the type of boundary conditions imposed.

For the deconfinement phase we have seen that the strong enhancement of
$D \bar{D}$ configurations earlier found for cooling with f.h.b.c. \cite{IMMPV1}
(which would be compatible with the suppression of the topological 
susceptibility) is not reproduced under cooling with standard p.b.c.
In the standard case, the probability to obtain any topologically
non-trivial object drops sharply with $\beta > \beta_c$. Cooling down to the
one-instanton action plateaus provides only trivial vacuum or
$M$-configurations. Because 
this latter observation was based on a physically small 3-volume 
($16^3 *4$ for $\beta =2.4$) finite-size effects might have been 
too strong to preclude
a final conclusion. The structure of cooled deconfined configurations
will be addressed in a further investigation.

The independence of the boundary conditions, however, in the confinement
phase has to be taken seriously: the enforcement of a $L=0$ boundary
condition seems to be not far from the situation with standard periodic
boundary conditions in the MC equilibrium. 
Some details will be discussed in the next Section.

\section{Dilute gas configurations of dyons and antidyons at higher action
plateaus}
\label{sect:gases}
Within the confinement phase, for $0 < T \le T_c$ and for both kinds of
spatial boundary conditions, we have also studied in detail semi-classical
configurations at higher action plateaus. 
They represent snapshots of earlier stages of the cooling histories
because the stopping criteria were focussed on multiples of the instanton
action. This study should allow us to observe superpositions of
classical solutions studied in Section \ref{sect:simple_lumps} 
promising to be relevant for 
a semi-classical approximation of the non-zero $T$ partition function. 
So far in the literature, the semi-classical approach to QCD at non-zero 
temperature is entirely based on Harrington-Shepard caloron solutions with 
trivial holonomy \cite{HS,GPY}. 
Our main concern is here, whether superpositions of solutions
with non-trivial holonomy naturally occur under cooling.

\subsection{Landscapes of topological density and Polyakov loop,
            fermion zero modes and Abelian monopoles as dyon finder}

Therefore, we expose equilibrium Monte Carlo lattice gauge field 
configurations to cooling,
this time stopping under criteria which apply to different, subsequent 
action windows. We have been monitoring the landscape of topological 
density, of the Polyakov line operator as well as
the localization of the fermionic zero-modes in the semiclassical candidate 
configurations. 

Searching for more complex approximate classical solutions we have 
modified our previous stopping criterion, triggering now on:
\begin{itemize}
\item 
$ (m-\frac{1}{2}) S_{\mathrm{inst}}~< S_n < 
     ~ (m+\frac{1}{2}) S_{\mathrm{inst}}, \qquad m = 2, 3, \cdots, \qquad$ and
\item $S_n-2~S_{n-1}+S_{n-2} < 0$.
\end{itemize}
In particular we inspected the first (highest) visible plateaus which
occurred at various $m$-values, 
typically in the range $m \simeq 8 \div 20$ for $\beta=2.2$ on a 
lattice of the size $16^3 \times 4$.
Then, we looked at the series of subsequent action plateaus. In terms of the
objects classified in Section \ref{sect:simple_lumps},
we have scanned the resulting plateau configurations.

For more definiteness concerning the moment of taking snapshots of the 
configurations undergoing cooling along a plateau, we have additionally 
introduced a measure $\Delta$ for the mean violation of the classical 
lattice field equations per link (see \cite{ILMPSS})
\eq
\Delta = \frac{1}{8 N_s^3 N_t} \sum_{x,\mu} \left\{
         \mathrm{tr} \left[ (U_{x,\mu}-\bar{U}_{x,\mu})
                            (U_{x,\mu}-\bar{U}_{x,\mu})^{\dagger}
                                            \right]\right\}^{\frac{1}{2}}\,,
\en
where
\eq \nonumber
\bar{U}_{x,\mu}= c \sum_{\nu > \mu} \left[  
U_{x,\nu}  U_{x+\hat{\nu},\mu} U^{\dagger}_{x+\hat{\mu},\nu} + 
U^{\dagger}_{x-\hat{\nu},\nu} U_{x-\hat{\nu},\mu} U_{x+\hat{\mu}-\hat{\nu},\nu}
                                    \right]
\en 
is the local link $x,\mu$ being the solution of the lattice equation of 
motion, with all degrees of freedom coupled to it being fixed. 
The factor $c$ is just a normalization of the staple sum such that 
$\bar{U}_{x,\mu} \in SU(2)$ \footnote{The replacement
$\bar{U}_{x,\mu} \rightarrow U_{x,\mu}$ is exactly 
the local cooling step as applied throughout this paper.}.

On the {\it first} visible plateau we find a gas of localized lumps
carrying topological charge, where an identification in terms of
dyons $D$ and/or antidyons $\bar{D}$ is still difficult.

\begin{figure}
\vspace{-3cm}
\begin{center}
\hspace{-0.9cm}$A$\hspace{0.2cm}$B$\hspace{2.6cm}$C$

($a$)\hspace{2.1cm}
\includegraphics[width=0.62\textwidth,height=0.40\textwidth]{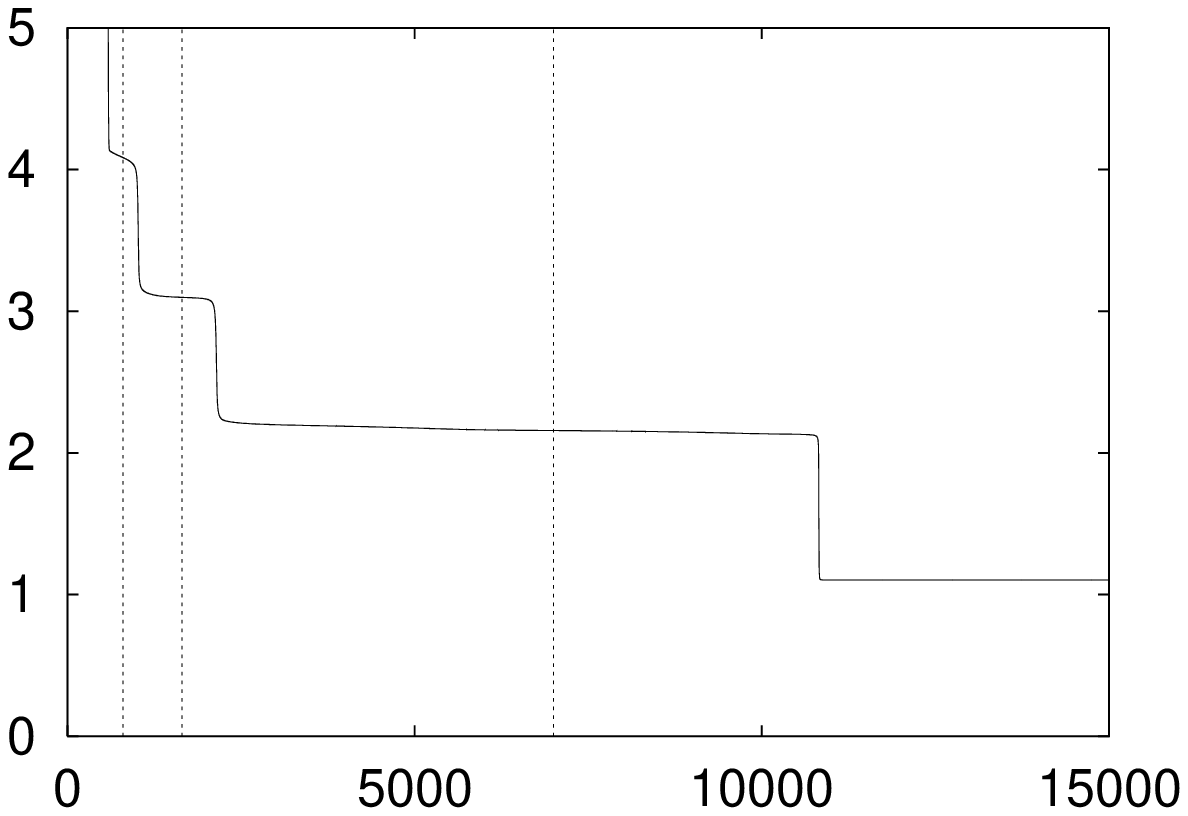}

\vspace{0.5cm}
($b$)\hspace{1.4cm}
\includegraphics[width=0.67\textwidth,height=0.40\textwidth]{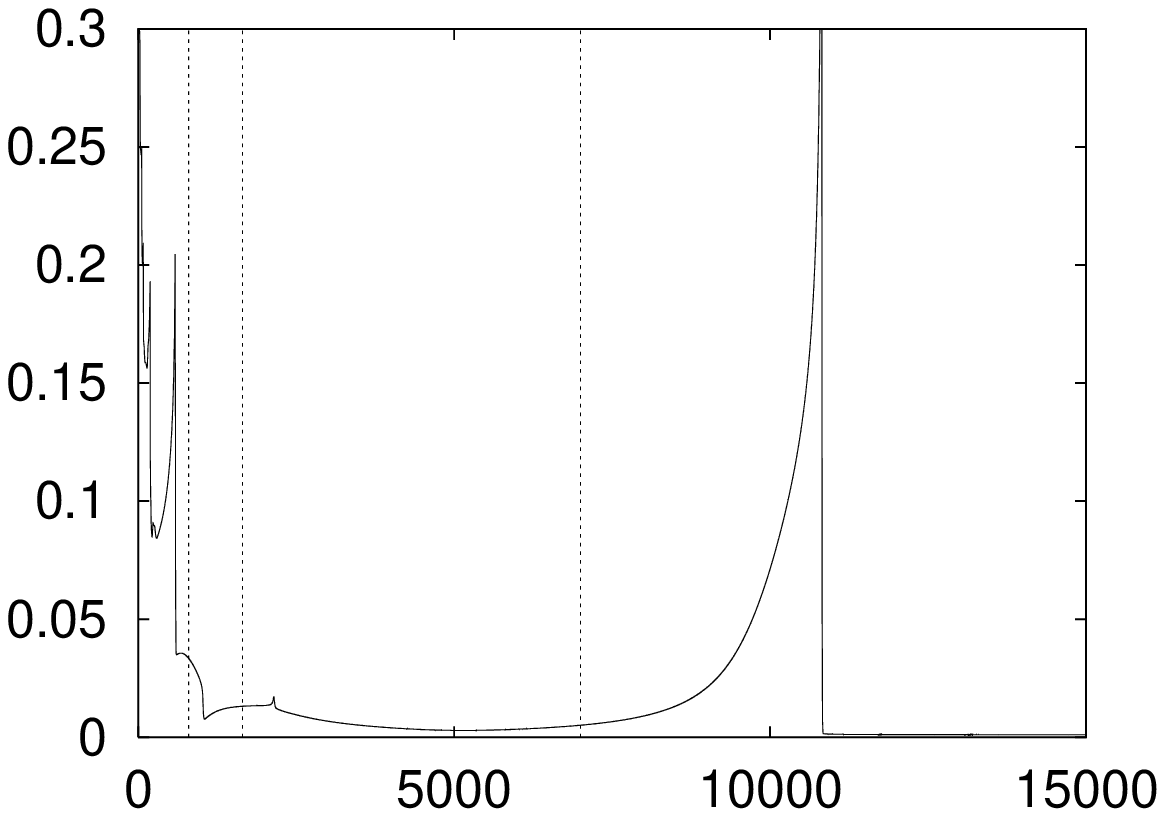}

\vspace{0.5cm}
($c$)\hspace{0.8cm}
\includegraphics[width=0.73\textwidth,height=0.40\textwidth]{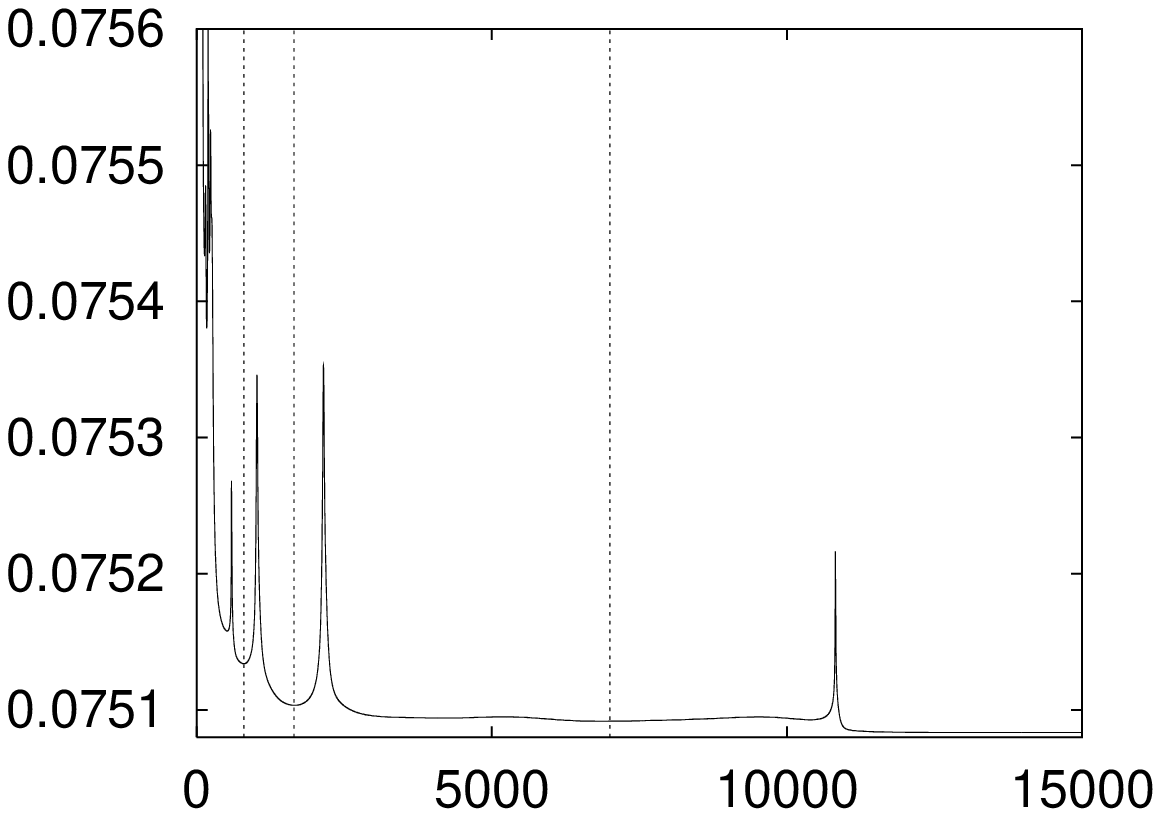}
\end{center}
\caption{
 Part of the cooling history for a gauge field configuration taken from 
 the Monte Carlo sample generated at $\beta=2.2$ on a $24^3 \times 4$ 
 lattice, with f.h.b.c. of $L_{\vec{x} \in \Omega}= 0$. 
 The sub-panels show:
 (a) full action $S/S_{inst}$,
 (b) non-stationarity $\delta_t$ and  
 (c) mean violation $\Delta$ per link of the lattice field equations,
 {\it vs.} the number of cooling steps.
 The vertical dotted   lines indicate the passages of $\Delta$ through local minima
 having occured at
 800 (A), 1650 (B) and 7000 (C) cooling steps for which the configurations
 will be portrayed in Figures \ref{fig:cool_q_pol}, \ref{fig:cool_fermion_1}, 
 \ref{fig:cool_fermion_2} and \ref{fig:cool_fermion_3}.}
\label{fig:history}
\end{figure}

\begin{figure}
\vspace{-2cm}
($A$)\hspace{-1.2cm}
\includegraphics[width=0.45\textwidth]{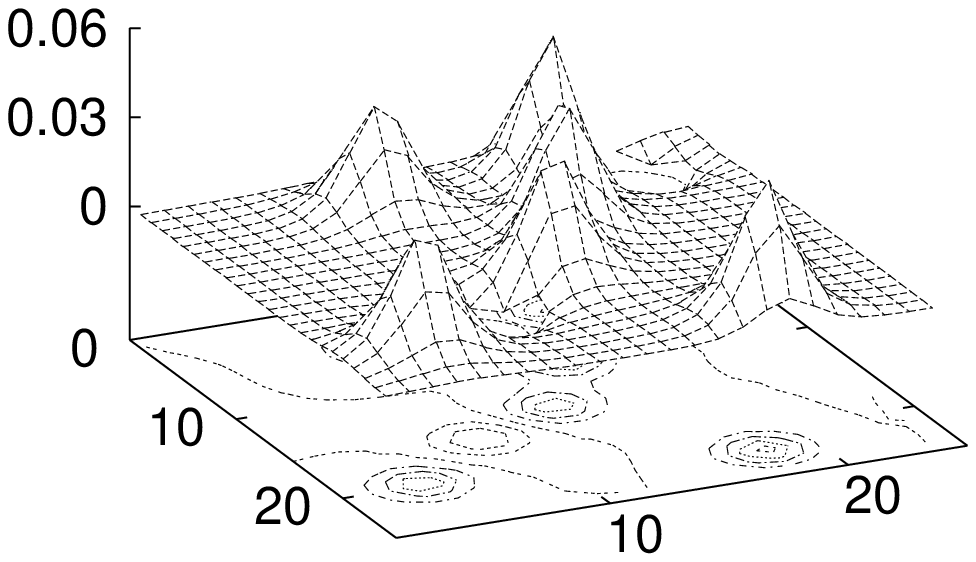}%
\hspace{0.1cm}($A'$)\hspace{0.2cm}
\includegraphics[width=0.45\textwidth]{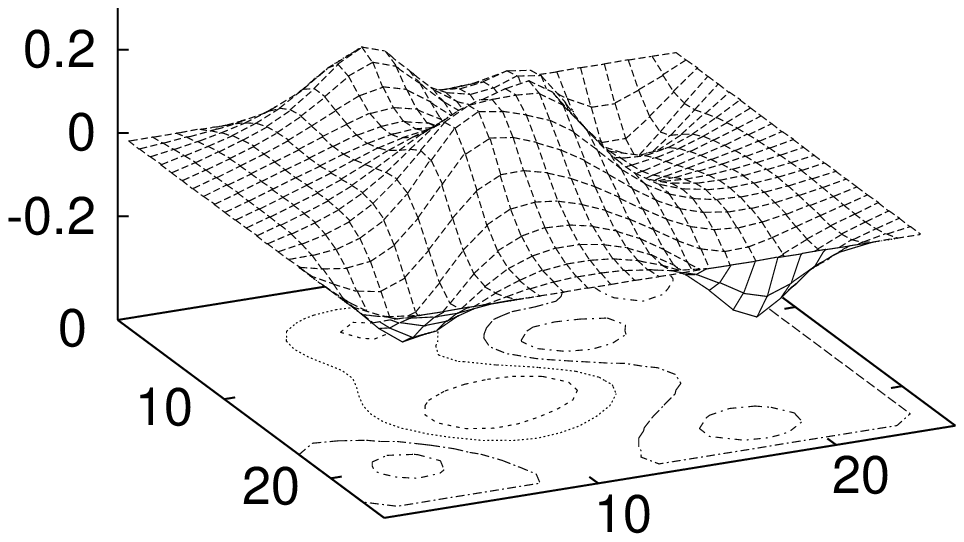} \\
\vspace{1cm}
($B$)\hspace{-1.2cm}
\includegraphics[width=0.45\textwidth]{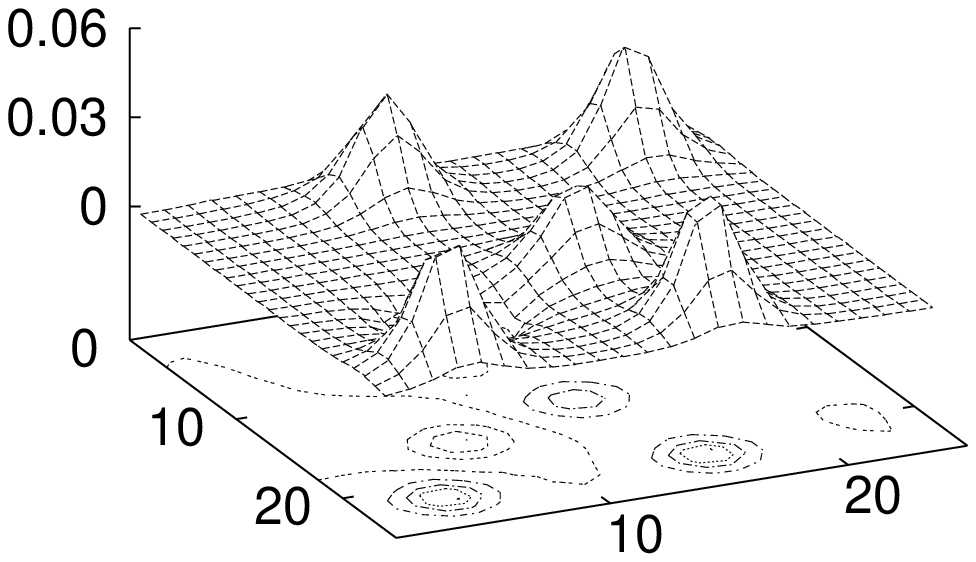}%
\hspace{0.1cm}($B'$)\hspace{0.2cm}
\includegraphics[width=0.45\textwidth]{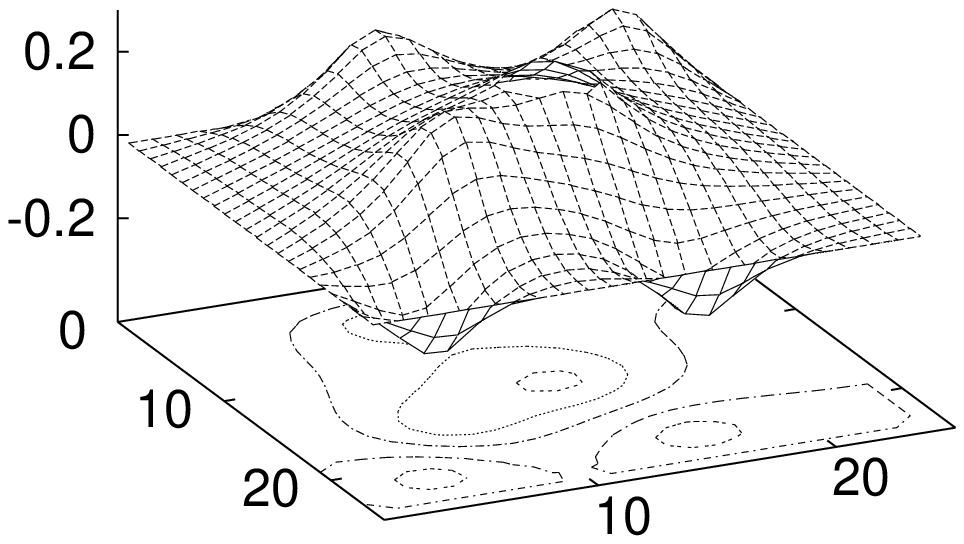} \\
\vspace{1cm}
($C$)\hspace{-1.2cm}
\includegraphics[width=0.45\textwidth]{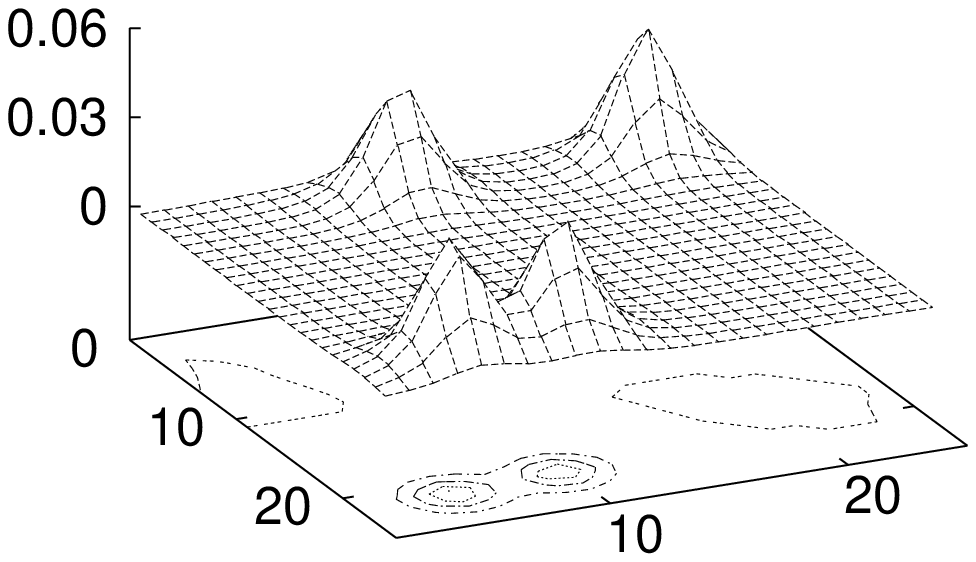}%
\hspace{0.1cm}($C'$)\hspace{0.2cm}
\includegraphics[width=0.45\textwidth]{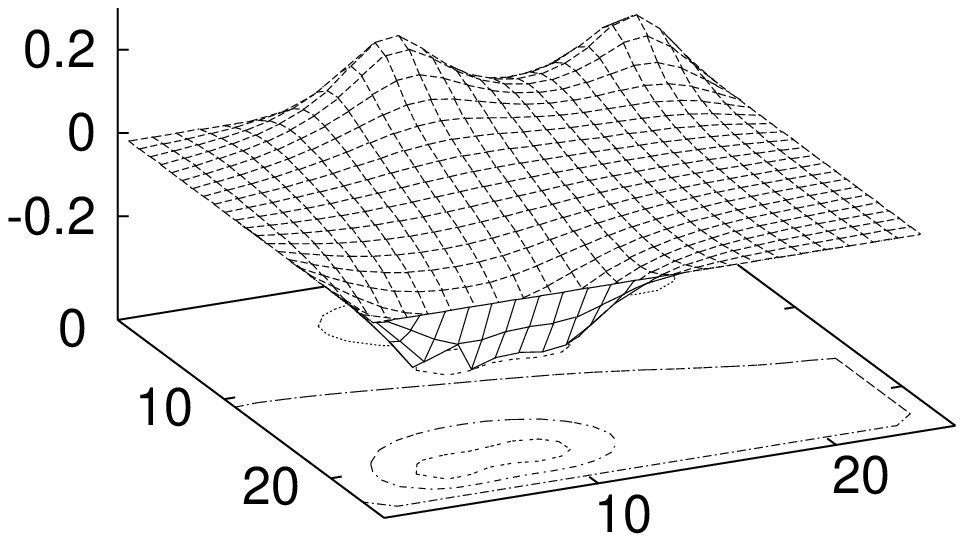}
\vspace{1cm}
\caption{
 Configurations on the $24^3 \times 4$ lattice (from equilibrium at 
 $\beta=2.2$ with f.h.b.c.), as indicated in Fig. \ref{fig:history}
 after 800 (A,A'), 1650 (B,B') and 7000 (C,C') cooling steps.
 (A,B,C) show $2D$ projections of the topological charge density $q_t(x)$ 
 and (A',B',C') of the Polyakov loop $L(\vec{x})$, respectively. 
 Cooling has been employed with f.h.b.c., $L_{\vec{x} \in \Omega}= 0$.}
\label{fig:cool_q_pol}
\end{figure}

\begin{figure}
\vspace{-3cm}
($A$)
\includegraphics[width=0.45\textwidth,height=0.35\textwidth]{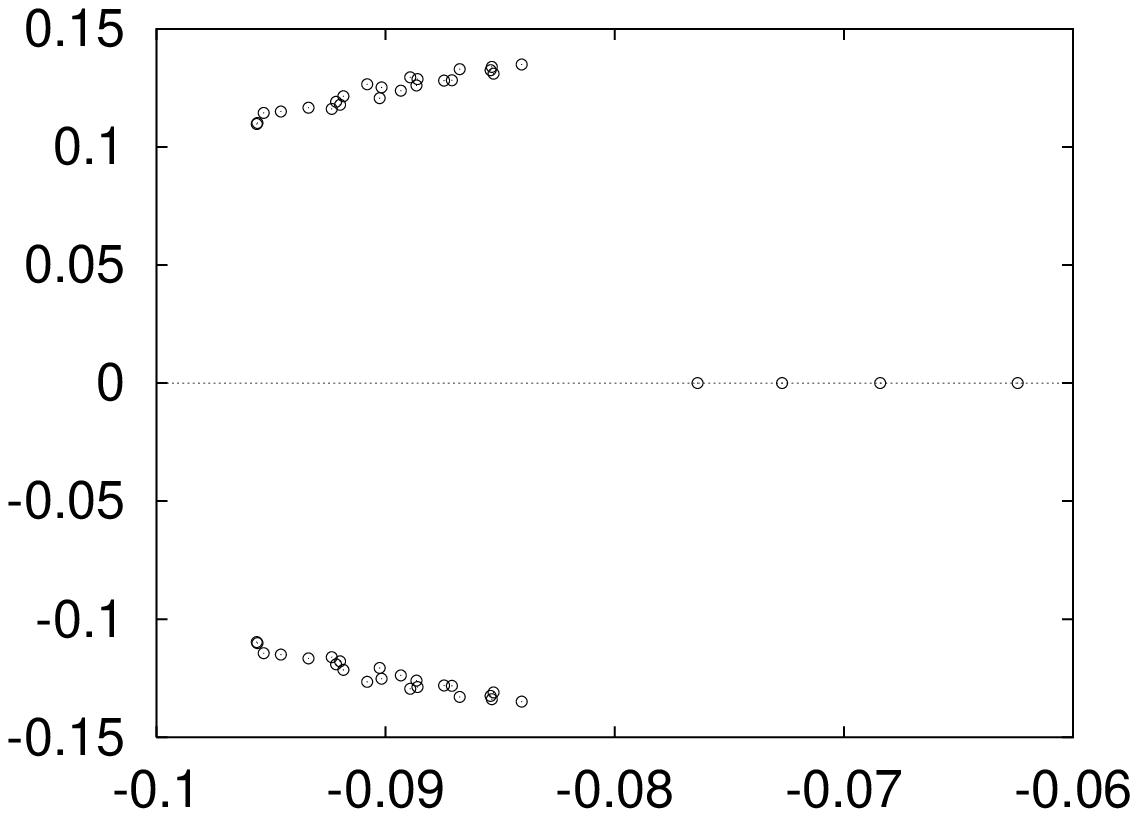}\\
\hspace{-0.4cm}($A1$)\vspace{0.2cm}
\includegraphics[width=0.45\textwidth,height=0.4\textwidth]{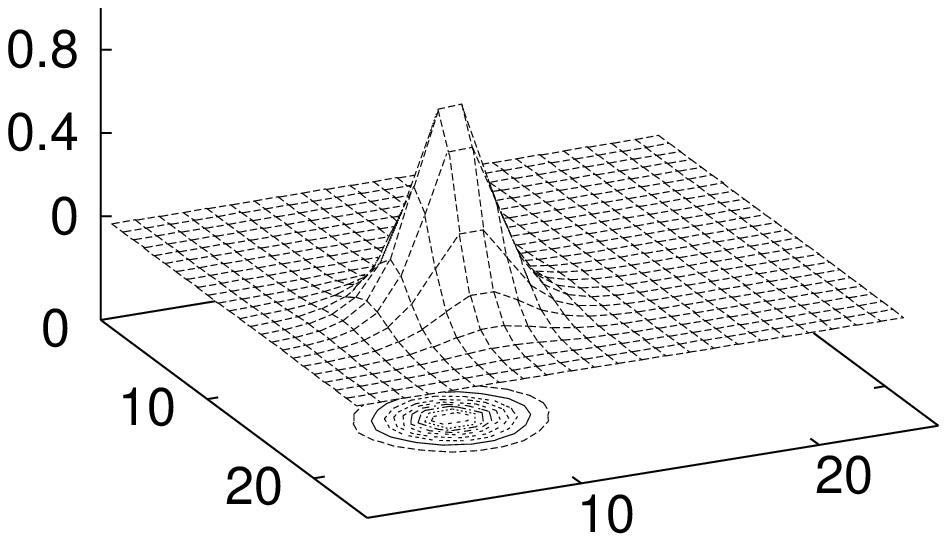}%
\hspace{-0.4cm}($A2$)\vspace{0.2cm}
\includegraphics[width=0.45\textwidth,height=0.4\textwidth]{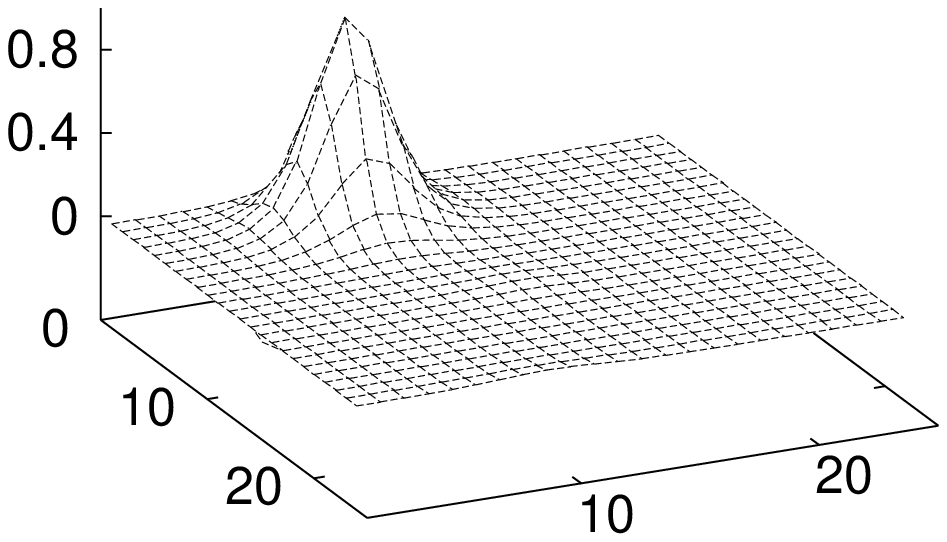}\\
\hspace{-0.1cm}($A3$)
\includegraphics[width=0.45\textwidth,height=0.4\textwidth]{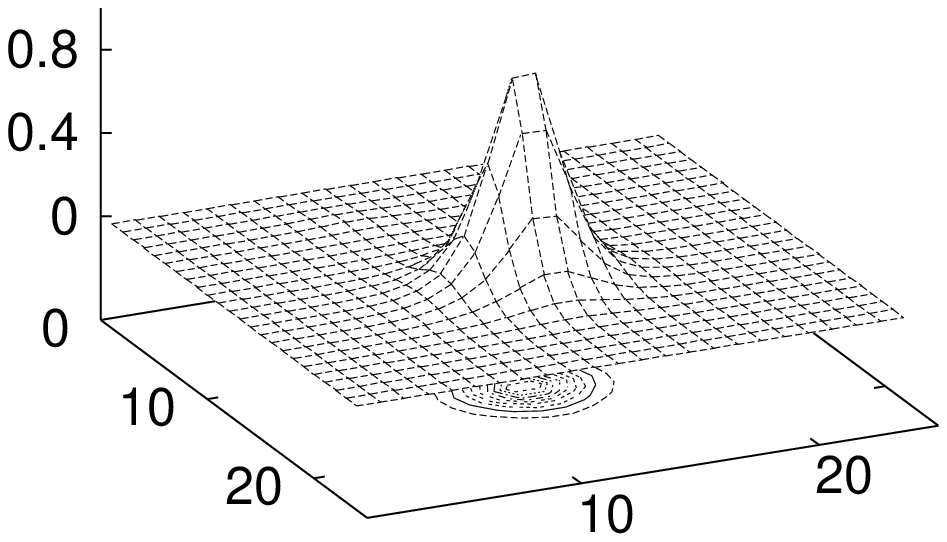}%
\hspace{-0.1cm}($A4$)
\includegraphics[width=0.45\textwidth,height=0.4\textwidth]{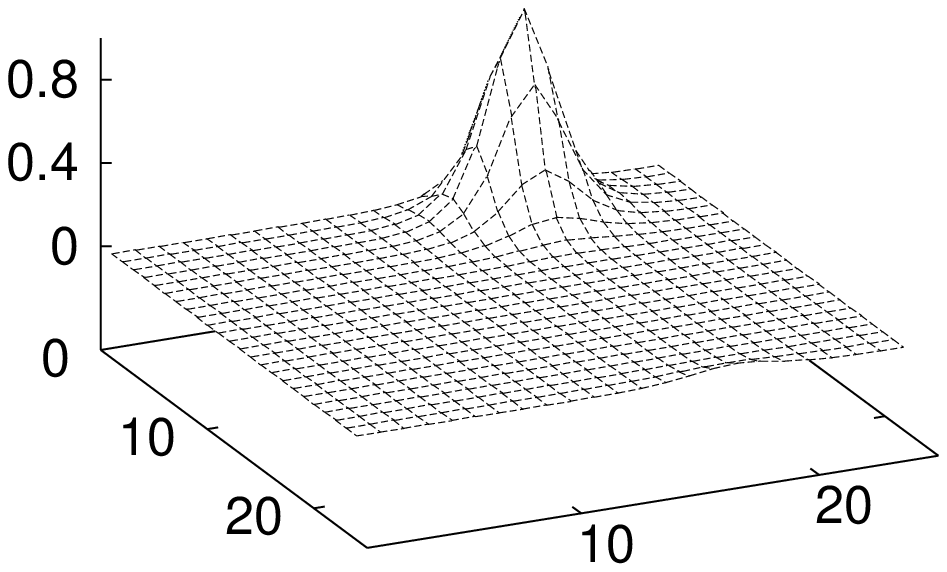}
\vspace{1.0cm}
\caption{ 
 The lattice field configuration depicted in Fig. \ref{fig:cool_q_pol}
 (A,A') for 800 cooling steps (f.h.b.c.). Here 
 (A) plots the eigenvalues of the Wilson-Dirac operator in the complex 
 plane for $\kappa=0.140$ and the case of time-periodic fermionic b.c.;
 (A1,...,A4) show $2D$ projections of the fermionic mode densities 
 related to the four distinct real eigenvalues.}
\label{fig:cool_fermion_1}
\end{figure}

\begin{figure}[!htb]
\vspace{-3cm}
($B$)
\includegraphics[width=0.45\textwidth,height=0.35\textwidth]{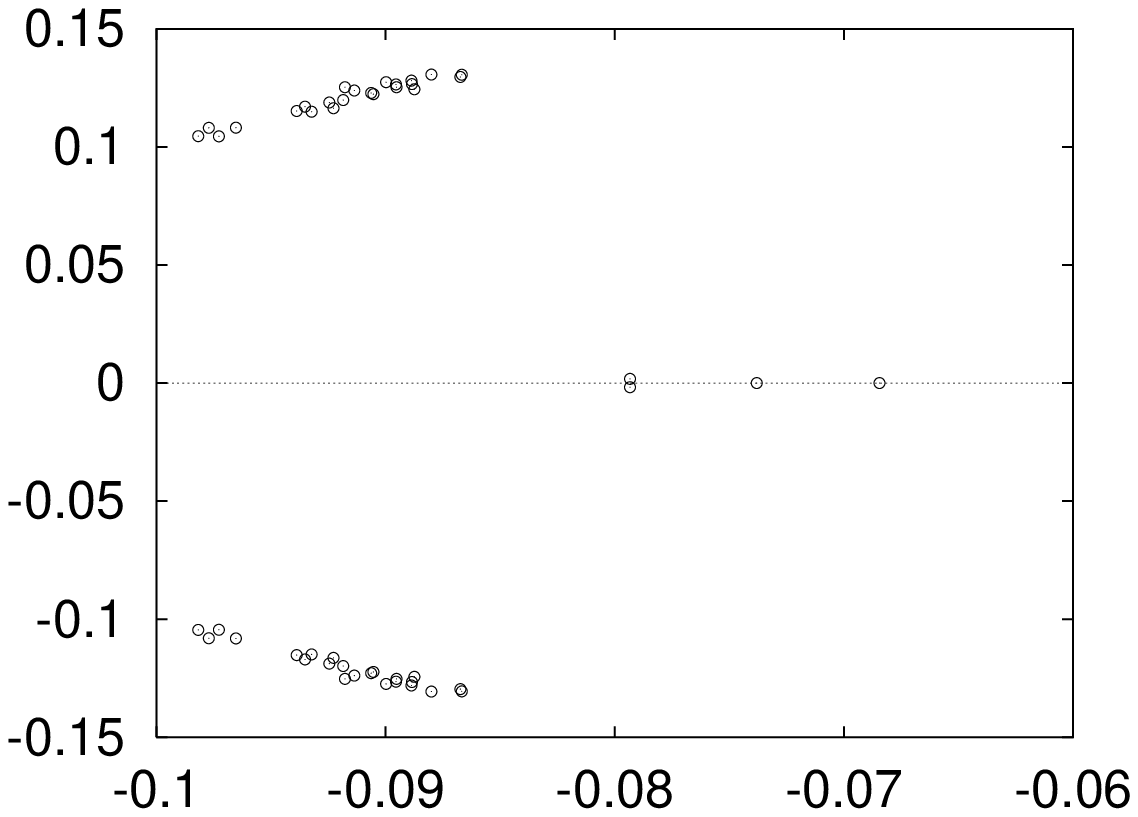}\\
\hspace{-0.4cm}($B1$)\vspace{0.2cm}
\includegraphics[width=0.45\textwidth,height=0.4\textwidth]{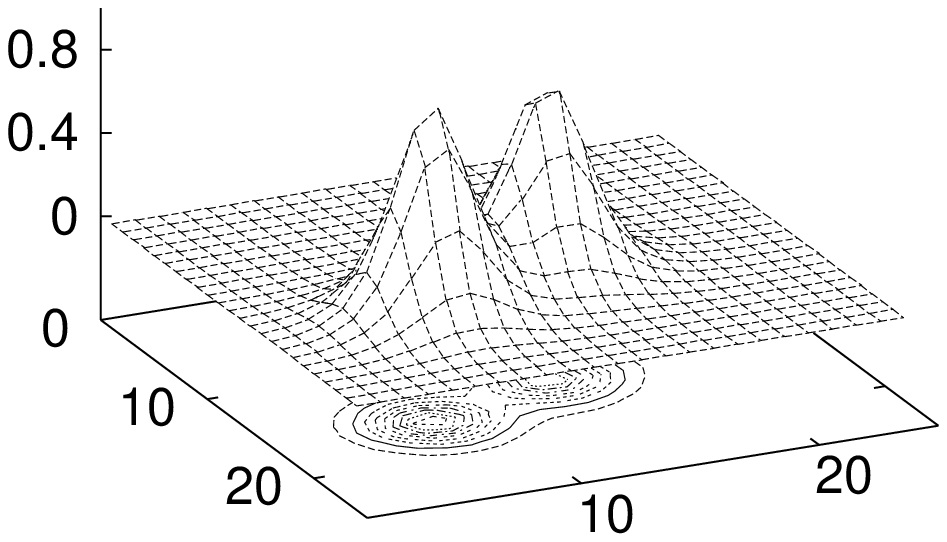}%
\hspace{-0.4cm}($B2$)\vspace{0.2cm}
\includegraphics[width=0.45\textwidth,height=0.4\textwidth]{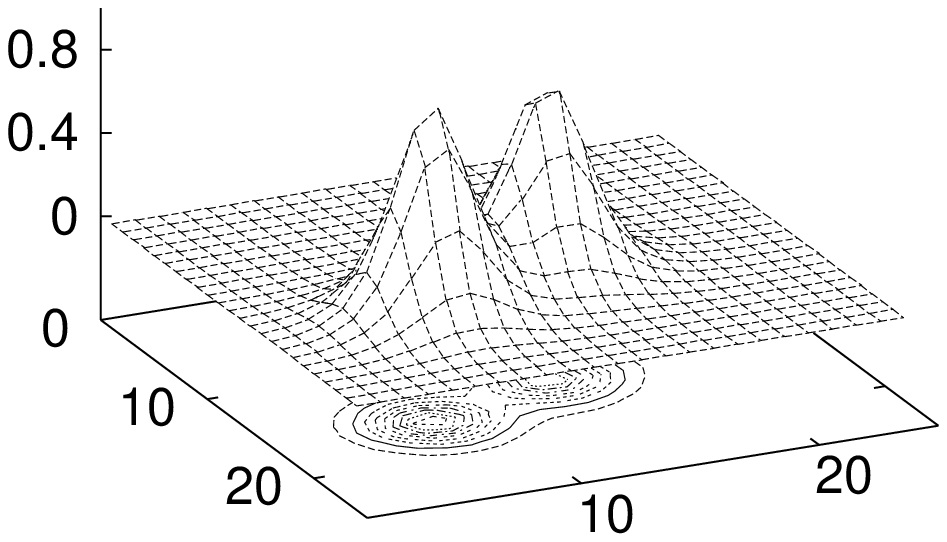}\\
\hspace{-0.1cm}($B3$)
\includegraphics[width=0.45\textwidth,height=0.4\textwidth]{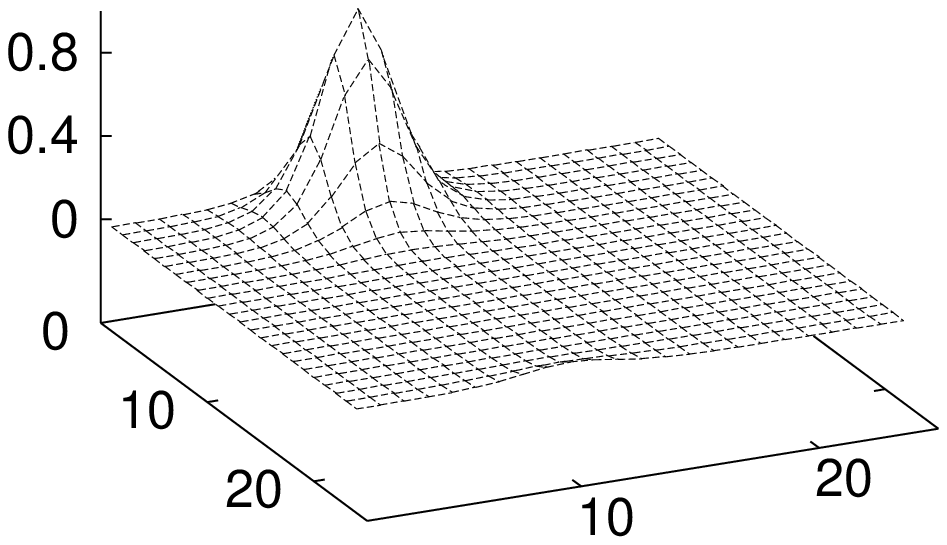}%
\hspace{-0.1cm}($B4$)
\includegraphics[width=0.45\textwidth,height=0.4\textwidth]{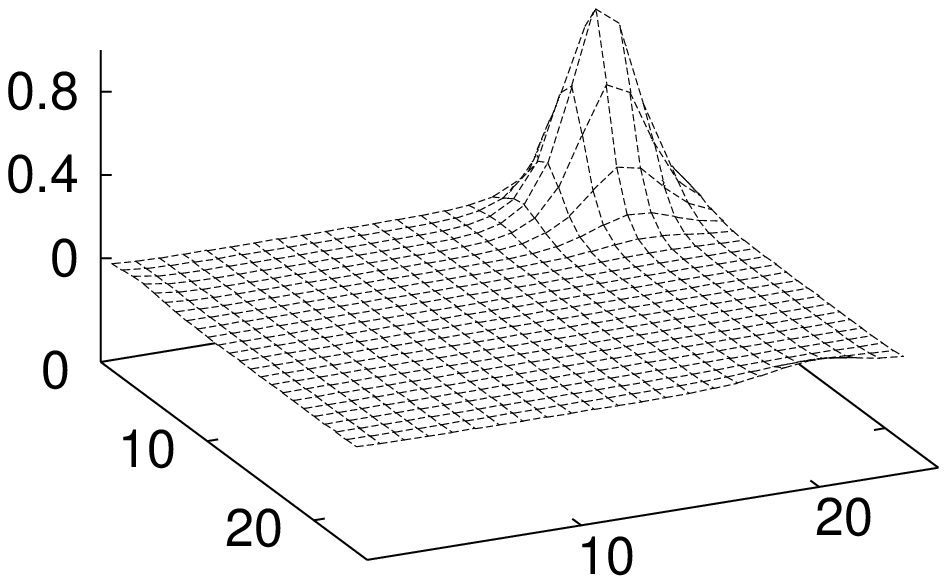}
\vspace{1.0cm}
\caption{
 The lattice field configuration depicted in Fig. \ref{fig:cool_q_pol}
 (B,B') for 1650 cooling steps (f.h.b.c.). Here 
 (B) plots the eigenvalues of the Wilson-Dirac operator in the complex 
 plane for $\kappa=0.140$ and the case of time-periodic fermionic b.c.;
 (B1,B2) show $2D$ projections of the fermionic mode densities related to
 the two distinct {\it almost} real eigenvalues, whereas (B3,B4) present
 the densities related to the two real eigenvalues.}
\label{fig:cool_fermion_2}
\end{figure}

\begin{figure}[!htb]
\vspace{-3cm}
($C$)\hspace{0.3cm}
\includegraphics[width=0.45\textwidth,height=0.35\textwidth]{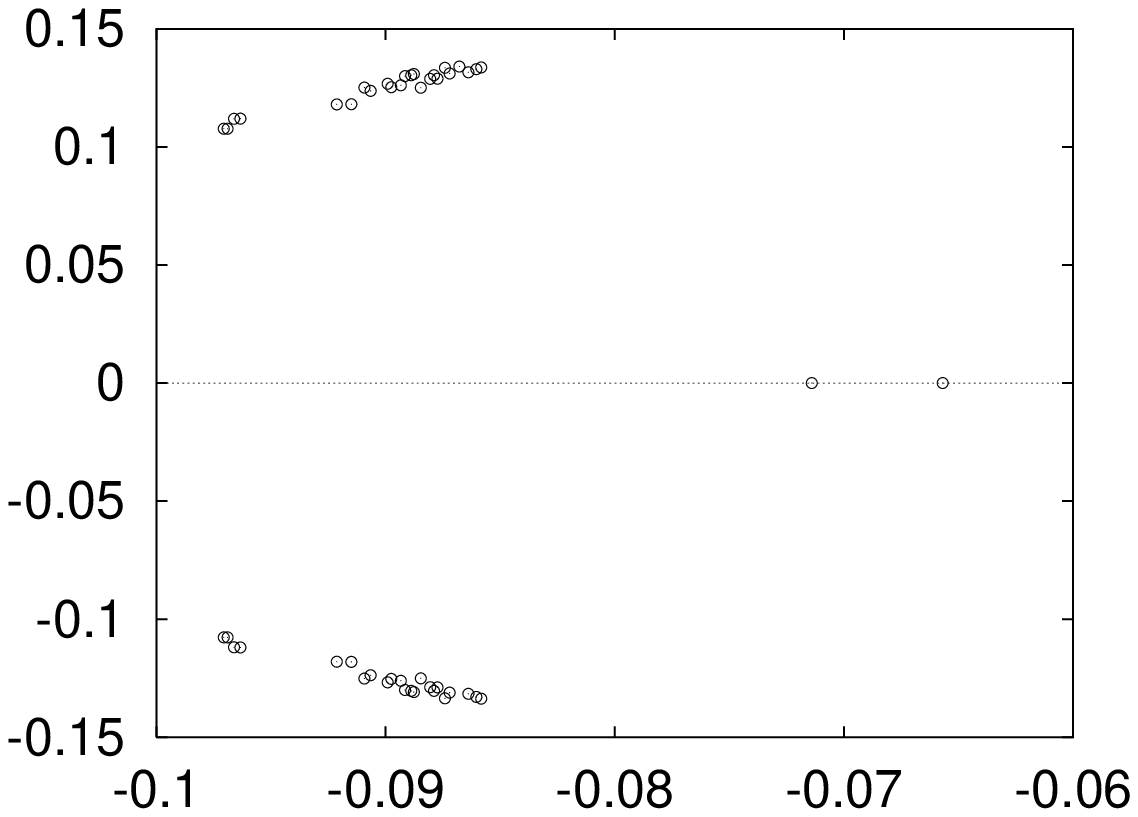}\\
\hspace{-0.4cm}($C1$)\vspace{0.2cm}
\includegraphics[width=0.45\textwidth,height=0.4\textwidth]{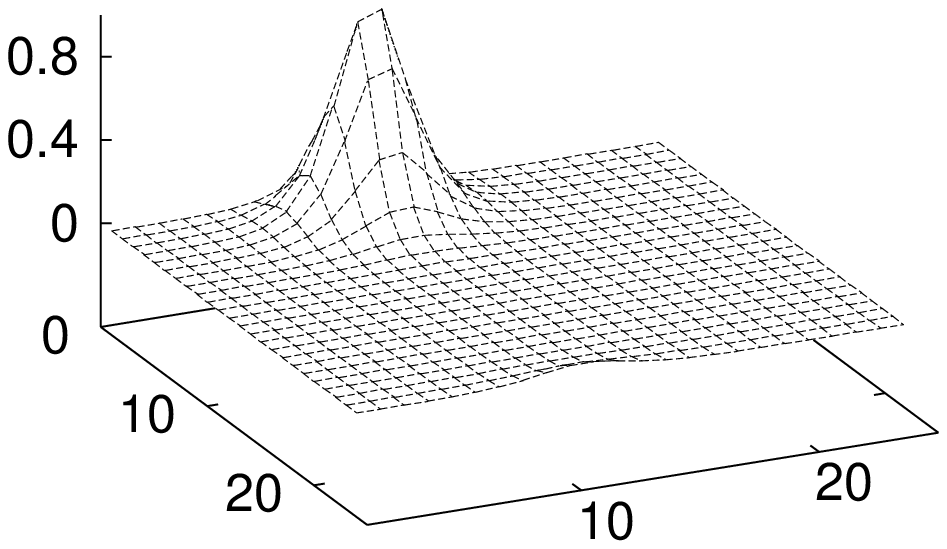}%
\hspace{-0.4cm}($C2$)\vspace{0.2cm}
\includegraphics[width=0.45\textwidth,height=0.4\textwidth]{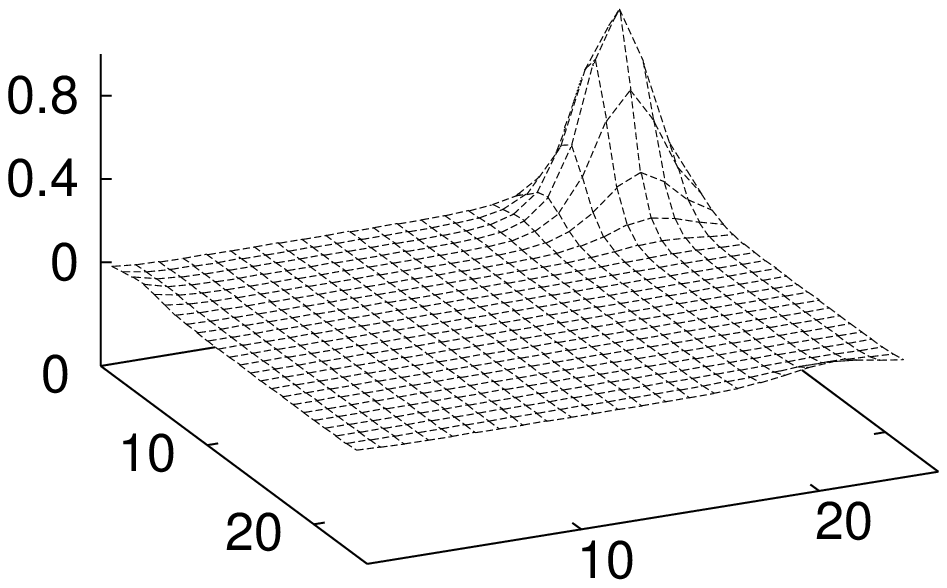}
\vspace{1.0cm}
\caption{
 The lattice field configuration depicted in Fig. \ref{fig:cool_q_pol}
 (C,C') for 7000 cooling steps (f.h.b.c.). Here 
 (C) plots the eigenvalues of the Wilson-Dirac operator in the complex 
 plane for $\kappa=0.140$ and the case of time-periodic fermionic b.c.;
 (C1, C2) show $2D$ projections of the fermionic mode densities related 
 to the two real eigenvalues.}
\label{fig:cool_fermion_3}
\end{figure}

Independent of the kind of boundary conditions employed,
at somewhat lower action plateaus with $m < 10$, we are able to clearly 
recognize dyons $D$ and antidyons $\bar{D}$ carrying 
non-integer topological charges.
During cooling more and more of these objects disappear. 
However, at all plateaus we observe an even number
of peaks of the spatial Polyakov loop landscape $|L(\vec{x})|$.
For illustration see Figs. \ref{fig:history} -- \ref{fig:cool_fermion_3}, 
which show one and the same gauge field configuration at different stages of
the cooling process. In this case f.h.b.c. have been used.

In Fig. \ref{fig:history} we show the action (in units of the instanton or 
caloron action), the non-stationarity $\delta_t$, 
and the mean violation $\Delta$ of the lattice field equations per
link. At three subsequent, already lower action plateaus (labelled by $m$)
we indicate the iteration steps 
A (for $m=4$), B (for $m=3$) and C (for $m=2$),
respectively, where $\Delta$ passes through
local minima. The corresponding semi-classical field configurations are then
displayed in Fig. \ref{fig:cool_q_pol} by means of the $2D$-projected
(by summing with respect to the third coordinate)
spatial topological charge density and the $2D$ projected Polyakov loop 
distribution.
More or less well one can recognize in these Figures that at stage A we have 
a superposition of 6 dyons and 2 antidyons. The topological charge sector 
has been independently determined to be $Q_t=2$. A $D \bar{D}$ pair decays  
or annihilates
from A to B such that we have 5 dyons and 1 antidyon at the next stage.
The topological charge did not change. Finally, stage C exhibits a superposition
of 4 dyons, again with $Q_t = 2$. The latter configuration is very stable.
While it stays at the same action over thousands of cooling sweeps, 
the non-stationarity $\delta_t$ gradually rises. 
A closer look then shows that the scale size of one of the dyon pairs 
shrinks, transforming this pair into a small undissociated and non-static 
caloron, which finally disappears after having turned into a tiny 
dislocation strongly violating the equations of motion 
(compare with Fig. \ref{fig:history}). 
The average violation of the equations of motion per link
peaks immediately before the configuration drops to the next plateau.

This example
shows that we have superpositions of non-integer $Q_t$ lumps, which can be
interpreted as described in the previous Section.  To make sure that this
is really the case we provide also the eigenvalue scatter plots for the
Wilson-Dirac operator for stage A (Fig. \ref{fig:cool_fermion_1}), for 
stage B (Fig. \ref{fig:cool_fermion_2}) and for stage C 
(Fig. \ref{fig:cool_fermion_3}).
Fig. \ref{fig:cool_fermion_1} shows 
four real fermion modes (under time-periodic boundary conditions) 
which could suggest an interpretation of 
configuration A as a superposition of 3 $DD$ pairs and a single 
$\bar{D} \bar{D}$ pair.
However, the inspection of the time-antiperiodic b.c. case provides only two
real modes which supports 
a dyonic content consisting of 2 $DD$ pairs plus 2 $D \bar{D}$
pairs, an interpretation which naively is possible as well.
In the stage C, also for time-periodic boundary conditions, we see clearly
two real modes sitting on top of two dyon lumps. We have checked that the
modes jump to the remaining dyon lumps when changing the fermionic boundary
condition to time-antiperiodic. 

\begin{figure}[!htb]
\vspace{-2cm}
\rotatebox{270}{\includegraphics[width=0.6\textwidth]{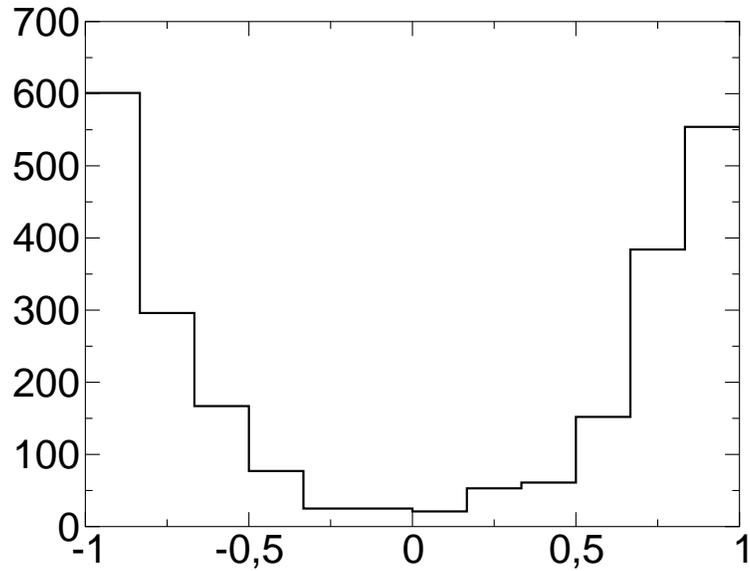}}
\caption{
 Histogram $P(L)$ of the values of the Polyakov loop $L(\vec{x})$ taken 
 at $\vec{x}$ where time-like Abelian (anti)monopoles are found.
 The data represent cooling plateaus at $m=4$
 obtained at $\beta=2.2$ with lattice size $16^3 \times 4$.
 O(2400) non-vanishing monopole charges were collected.
}
\label{fig:corr_mon_pol}
\end{figure}

We have studied also the Abelian (anti)monopole structure after fixing 
to the MAG and Abelian projection.
We see a strong correlation of the peaks of the Polyakov loop with the
positions of the (anti)monopoles. 
This can be seen in Fig. \ref{fig:corr_mon_pol}.
The pair structure in terms of Abelian monopoles, occuring on all action 
plateaus and the annihilation of single (monopole-antimonopole) pairs 
during further relaxation, provides an additional
signal for the topological content 
as superpositions of non-trivial holonomy $CAL$, $DD$ ($\bar{D} \bar{D}$) 
or $D \bar{D}$ pairs.

\subsection{The r\^ole of the boundary conditions: 
            are non-trivial boundary conditions ''natural'' ?}

In order to understand this from the point of view of single caloron
solutions with non-trivial holonomy we have to find out 
whether (approximately) those asymptotic holonomy boundary conditions as 
typical for the dyonic (antidyonic) semi-classical background excitations 
are actually present 
during the cooling process when periodic boundary conditions
(no particular holonomy boundary conditions !) 
are applied to the full volume. 
Then it would be easier to accept that similar (albeit fluctuating)
boundary conditions might also be realized in the semi-classical vacuum.

To answer this question we {\it define} the {\it asymptotic holonomy} 
$L_{\infty}$ of a cooled configuration as the average of $L(\vec{x})$ 
over all points $\vec{x}$ in $3D$ space for which the 
$3D$ projected action
density satisfies ${\varsigma}(\vec{x}) \le .0001$, i.e. it takes minimal
values which are typically seen in deep valleys around the 
topological lumps.

\begin{figure}[!htb]
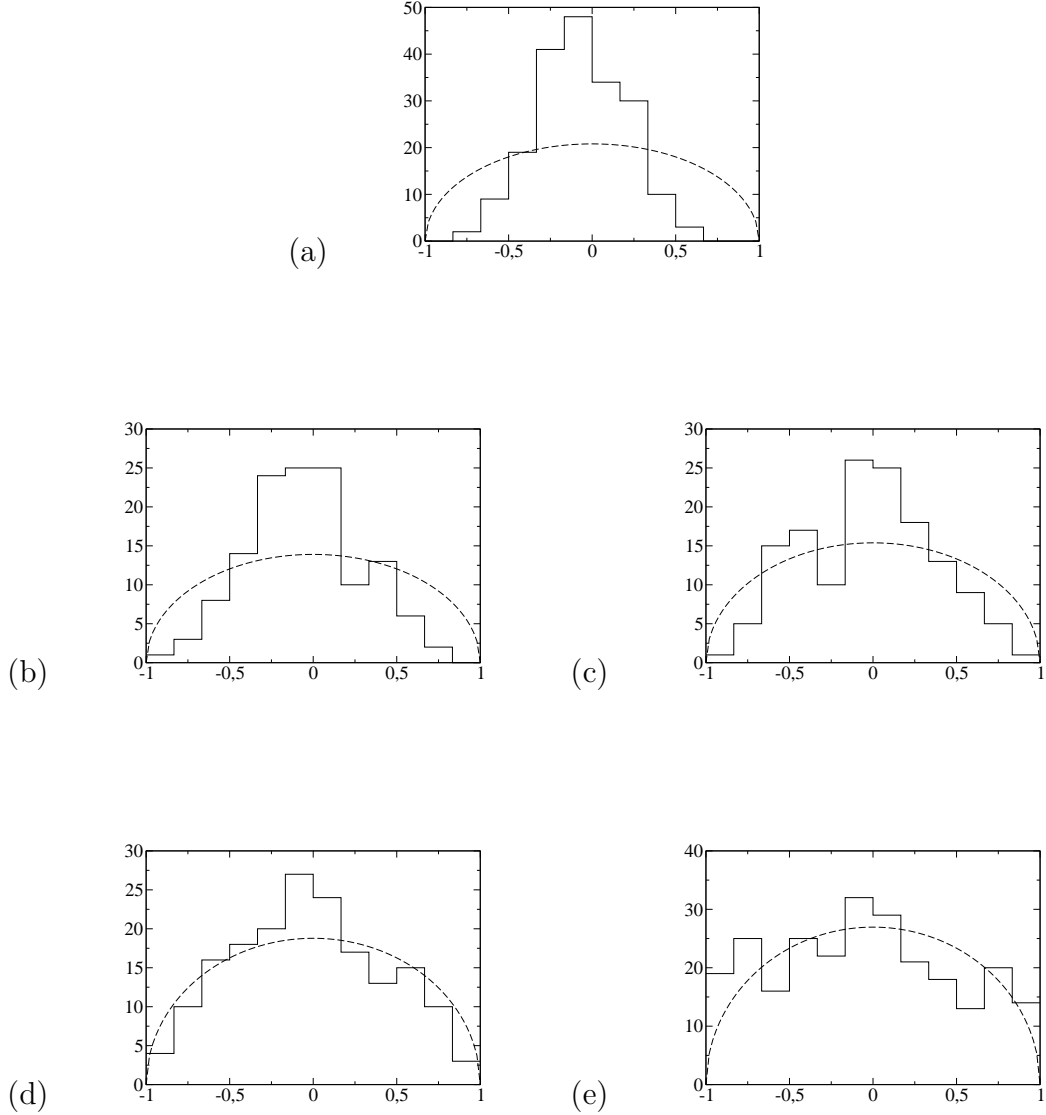

\begin{center}
(a)\hspace{1cm}\includegraphics[width=0.3\textwidth]{hol.eps}

\vspace{2cm}
(b)\hspace{1cm}\includegraphics[width=0.3\textwidth]{hol4.eps}%
\hspace{1cm}
(c)\hspace{1cm}\includegraphics[width=0.3\textwidth]{hol3.eps}

\vspace{2cm}
(d)\hspace{1cm}\includegraphics[width=0.3\textwidth]{hol2.eps}%
\hspace{1cm}
(e)\hspace{1cm}\includegraphics[width=0.3\textwidth]{hol1.eps}
\end{center}
\vspace{2cm}
\caption{
 Histogram $P(L_{\infty})$ of the values of the Polyakov loop 
 at ''infinity'' (as explained in the text) 
 seen on the first plateau (a) and at plateaus with $m \simeq 4$ (b), 
 $m \simeq 3$ (c), $m \simeq 2$ (d), $m \simeq 1$ (e).
 For comparison the distribution expected from the pure Haar measure
 $P_{\rm Haar}(L) \sim \sqrt{1-L^2}$ 
 is shown with the same normalization (dashed lines). 
 The equilibrium ensemble was generated at $\beta=2.2$, 
 the lattice size is $16^3 \times 4$, cooling was performed using 
 periodic b.c., O(200) configurations were investigated.
}
\label{fig:holonomy_dist}
\end{figure}

In Fig. \ref{fig:holonomy_dist} we show histograms of $L_{\infty}$
obtained at different plateaus during the cooling histories of an ensemble
of $O(200)$ configurations produced at $\beta =2.2$ on a $16^3 \times 4$
lattice with standard p.b.c. We see a clear peak at $L_{\infty}=0$ for higher
action plateaus. 
The distribution is more narrow, than the pure Haar measure would tell us.
However, approaching lower-lying plateaus, the real distribution becomes 
flat.  
 
\begin{figure}[!htb]
\begin{center}
(a)\includegraphics[width=0.4\textwidth]{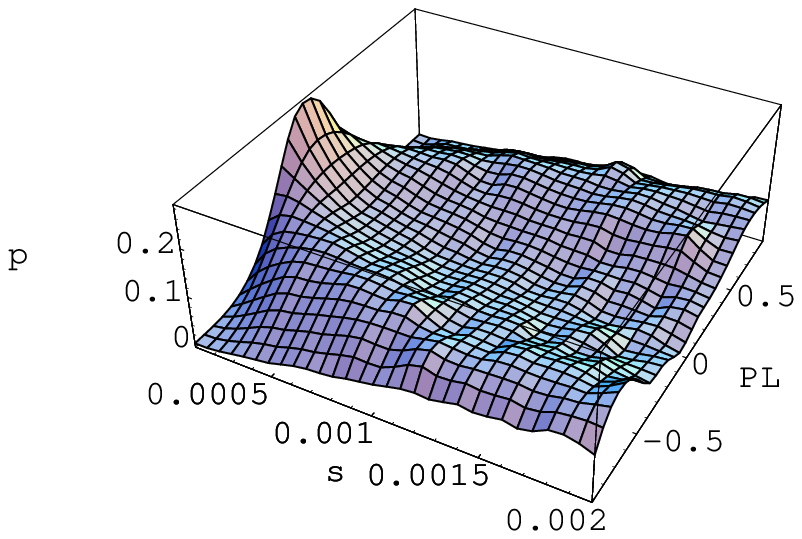}%
(b)\includegraphics[width=0.4\textwidth]{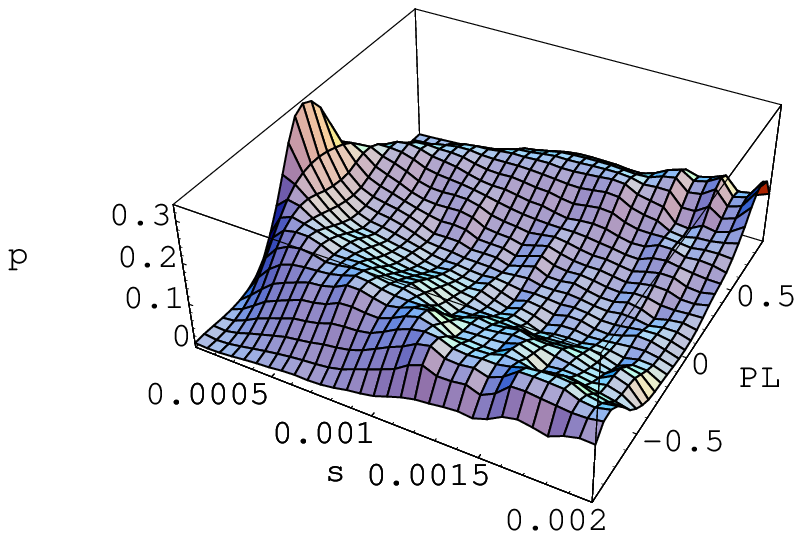}
(c)\includegraphics[width=0.4\textwidth]{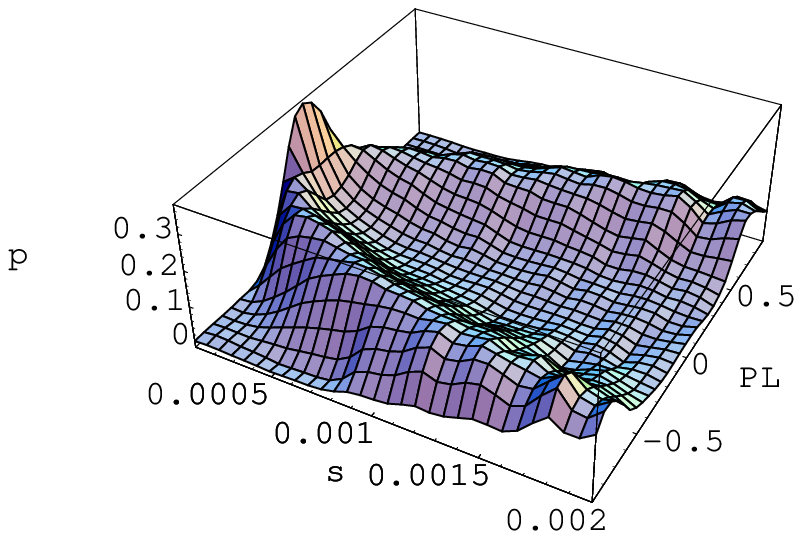}%
(d)\includegraphics[width=0.4\textwidth]{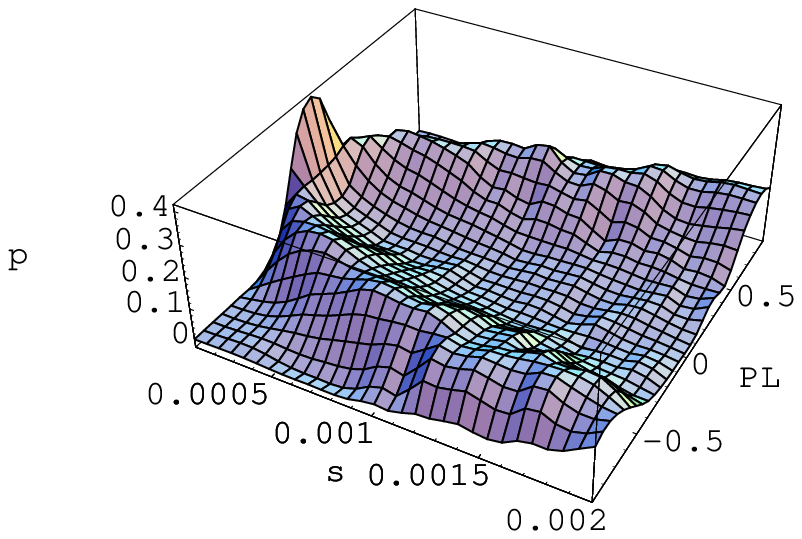}
\end{center}
\caption{
 Conditional distributions $P[L|\varsigma]$ relating local values
 of $L(\vec{x})$ with the spatial action density $\varsigma(\vec{x})$ 
 for cooled configurations at plateaus with $7 \ge m \ge 5$ (a), 
 $m \simeq 4$ (b), $m \simeq 3$ (c) and $m \simeq 2$ (d). 
 The equilibrium ensemble was generated at $\beta=2.2$, 
 the lattice size is $16^3 \times 4$, cooling was performed using p.b.c.}
\label{fig:cond_dist}
\end{figure}

\begin{figure}[!htb]
(a) \includegraphics[width=0.4\textwidth]{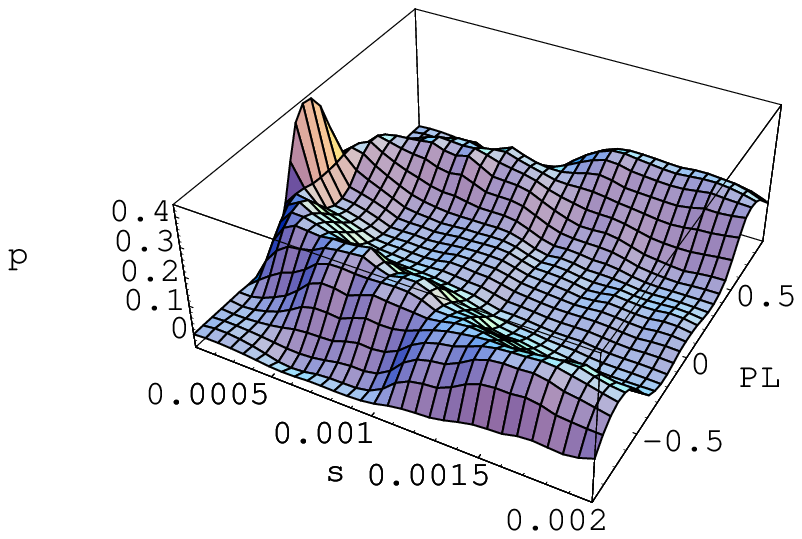}
(b) \includegraphics[width=0.4\textwidth]{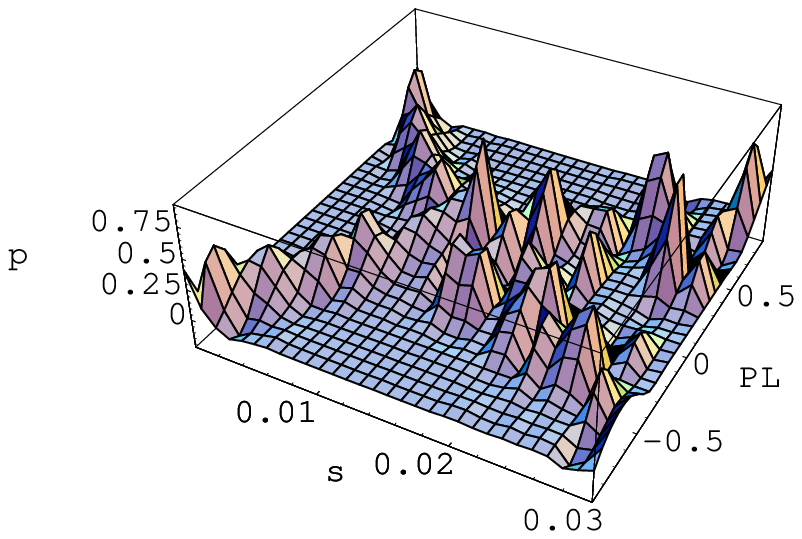}
\caption{
 Conditional distributions $P[L|\varsigma]$ as in Fig. \ref{fig:cond_dist},
 obtained for random KvB solutions ($DD$ or $CAL$) (a) and for calorons 
 with trivial holonomy (b), for comparison.}
\label{fig:cond_dist_vB}
\end{figure}

Closing the discussion of the local boundary conditions,
let us finally concentrate on those configurations which belong to the
bins of $-1/6 < L_{\infty} < +1/6$, i.e. those which realize 
$L_{\infty} \approx \langle L \rangle$ in the confined phase. 
For these configurations we turn our attention
to the local correlation between
the Polyakov loop $L(\vec{x})$ and the action density values
${\varsigma}(\vec{x})$ measured at the same positions $\vec{x}$.
We studied {\it conditional distributions} $P[L|\varsigma]$
of the Polyakov loop values $L(\vec{x})$ at 
spatial points where the spatial action density $\varsigma(\vec{x})$
equals $\varsigma$. The distributions are normalized for each bin of 
$\varsigma$. The distributions 
are shown as surface plots in Fig. \ref{fig:cond_dist} for different
plateaus. They show that with higher 
action density the corresponding
local Polyakov loop values 
tend to be closer to the peak values $L = \pm 1$. 
The distributions do not depend on which action plateau they were collected.
We have identified in Section \ref{sect:simple_lumps} 
KvB solutions and $D \bar{D}$ configurations
at the one-instanton action plateau, and we have strong indications that the
same objects occur on the higher action plateaus, too, i.e. in superpositions
of topologically non-trivial lumps of action. This argument is 
also supported by a comparison with the same kind of conditional distribution
obtained from the analytic KvB solution with random parameter
distribution which we have discretized on the lattice. The resulting 
distribution $P[L|\varsigma]$ is shown in 
Fig. \ref{fig:cond_dist_vB} (a) and compared with the distribution for
calorons with trivial holonomy (b). 

\subsection{The Taubes rotation in many-dyon configurations}

Finally, it is interesting to analyse the relative orientation
of the dyons in color space. In Section \ref{sect:formulae} 
we have described how the Taubes winding
in a $DD$ caloron could be detected in a gauge independent way.
The analysis is also here, for real cooled configurations, 
simplified by the observation
that at the center of a dyon both electric $E^a_i$ and magnetic $B^a_i$ field
strengths (which satisfy (anti) self-duality $E^a_i = \pm B^a_i$ ) form
orthogonal matrices in color ($a$) and space ($i$) indices.
Thus, it suffices to consider three ($i=1,2,3$) vectors in color space 
$E^a_i(\vec{x})$ forming a local reference frame at the center of a dyon
with which any other local reference frame can be compared.  The comparison
can be made in a gauge invariant manner by connecting the centers at which
the field strengths are measured by a parallel transporter, the Schwinger line.

The exploration of a few of semi-classical lattice configurations containing 
superpositions of several $D$ and $\bar{D}$ has shown the following
common features.
All dyons with negative peak value of the Polyakov loop have 
more or less random color
orientation relative to each other, but this relative orientation is static 
along the time axis.
For dyons with a positive peak value of the Polyakov loop holds the same.
Also their orientation relative to each other seems to be
random but static.
But the relative orientation between the $L=+1$ and the $L=-1$ dyons
is changing along the Euclidean time and the change is nothing but a 
homogeneous rotation in color space with the angle period $2\pi$.
In the analytic solution representing just two dyons (or two antidyons)
this rotation is also present. It has been discussed and 
related to Taubes winding in Sect. \ref{sect:formulae}.

To illustrate this observation let us consider a configuration obtained
on the lattice of size $16^3 \times 4$ with f.h.b.c. after 400 cooling steps.
The configuration contains three $D$ and one $\bar{D}$ as shown in Table 2.
\begin{table}[ht]
\caption{Configuration of 3 $D$ and 1 $\bar{D}$, obtained after 400
cooling steps with f.h.b.c. from an equilibrium configuration
produced at $\beta=2.2$ on a $16^3 \times 4$ lattice.}
\vspace{5mm}
\begin{center}
\begin{tabular}{lcc}
\hline
Type                &  $3D$ position            &  Polyakov loop  \\
\hline
$D$                 & $\vec{x}_1=(~5,~3,~2) $   &            $ -1 $          \\
$D$                 & $\vec{x}_2=(12,~6,~3) $   &            $ -1 $          \\
$D$                 & $\vec{x}_3=(13,14,~5) $   &            $ +1 $          \\
$\bar{D}$           & $\vec{x}_4=(~5,~7,13) $   &            $ -1 $          \\
\hline
\end{tabular}
\end{center}
\end{table}
We have taken 
the first $D$ as the reference point with
respect to which the relative 
color orientations of the other dyons and the antidyon
were determined. We have calculated the matrices (cf. eq. (\ref{eq:Rotation}))
$R^{12}_{ik}(t)$, $R^{13}_{ik}(t)$ and $R^{14}_{ik}(t)$.  
They show the relative orientations of the objects $n=2,3,4$ -- represented by
their electric fields $E_k(\vec{x}_n,t)=E^a_k(\vec{x}_n)\frac{\tau^a}{2}$ 
which appear parallel transported to the position of the first object 
in the form ${\cal S}(\vec{x}_1,t;\vec{x}_n,t)~E_k(\vec{x}_n,t)
~{\cal S}^{+}(\vec{x}_1,t;\vec{x}_n,t)$ --
with respect to the first object -- represented by its electric fields
$E_k(\vec{x}_1)$.  
Then the evolution in time of the relative orientation can be investigated.
While the matrices
$R^{12}_{ik}(t)=\mathrm{tr} \left( E_i(\vec{x}_1,t) {\cal S} 
                                    E_k(\vec{x}_2,t) {\cal S}^{+} \right)$,
and $R^{14}_{ik}(t)$ turned out to be constant in time, 
the matrix $R^{13}_{ik}(t)$
performed a color rotation with a constant  
angle increment $\pi/2$ from one time slice 
to the next time slice about the color axis $\vec{n}=(-0.919,0.278,0.275)$. 
The orientation of this axis seems to be
random, but the rotation angle is well-defined. 
For $N_t=4$ it corresponds to a full
color rotation over the full Euclidean time period. 
Our general observation illustrated 
by this example of a moderately complicated
superposition shows that also these more complicated objects
exhibit a strong correlation in the color orientation analogous 
to that present in a single $DD$ KvB pair. 
A semi-classical approximation of the path integral would have
to take into account this kind of color correlation.

\section{Conclusions}
\label{sect:conclusions}
We have generated $SU(2)$ lattice gauge fields at non-zero temperature 
in the confinement phase. We have cooled them in order to 
analyse their topological content. Fixed holonomy spatial boundary
conditions have been used as well as standard periodic boundary conditions.
The results for these two kinds of boundary conditions semi-quantitatively
agree with each other. This is specific for ensembles describing the 
confinement phase.

Independently of the boundary conditions we have found superpositions of
calorons, dyons and antidyons, the latter with positive and negative
non-integer topological charges. The topological lumps appear also 
as peaks of the Polyakov loop modulus $|L(\vec{x})|$, with calorons
being a limiting case with a close pair of 
$L(\vec{x})=\pm 1$.
Investigating also the localization behavior of the real eigenvalue
modes of the (non-Hermitean) Wilson-Dirac operator we could present
convincing evidence that for calorons and for dyon-dyon pairs an 
interpretation in terms of KvB solutions 
is natural. Chosing time-periodic and time-antiperiodic boundary conditions
for the fermions focusses on $L(\vec{x})=-1$ or $L(\vec{x})=1$ dyons,
respectively.

On higher action plateaus we have found that the dynamics generically leads
to non-trivial holonomy outside the lumps of action and topological
charge. The multi-dyon-antidyon structure can be finally resolved by a
combination of two tools: localization of fermionic real modes
and the relative color-orientation of the color-electric field strength.
 
A semi-classical treatment of the path integral at non-zero temperature
close to the deconfinement phase transition should be built on superpositions
of calorons, dyons and antidyons, however with the holonomy as a free parameter.

We have seen that such superpositions would imply 
a strong correlation in the relative color orientation (and its Euclidean
time dependence) between pairs of seemingly independent topological lumps. 
To the best of our knowledge,
such superpositions have not yet been constructed analytically.

It is already clear that the development of a semi-classical approach based 
on solutions with non-trivial holonomy is much more complicated than
the instanton (caloron) gas or liquid, and it might turn out not to lead to
a practicable model. 

Nevertheless, facing the non-trivial structures found in this paper might 
contribute to a better understanding of
the mechanism driving the deconfinement transition. In as far a certain 
working picture of a dilute gas of these configurations can be developed
and whether it will improve our understanding of quark confinement itself 
remains an open question to which we hope to come back in the next future.

\section*{Acknowledgments}
We are grateful to P. van Baal and Z. Fodor for discussions and
useful comments. One of us (S.S.) acknowledges computational assistence 
by F. Hofheinz.
E.-M. I. gratefully appreciates the support by the Ministry
of Education, Culture and Science of Japan (Monbu-Kagaku-sho) 
and thanks H. Toki for the hospitality at RCNP.

This work was partly supported by RFBR grants 02-02-17308 and 01-02-17456, 
by the INTAS grant 00-00111 and the CRDF award RP1-2364-MO-02.


%
\newpage
\newpage
\newpage
\newpage
\newpage
\newpage
\newpage
\newpage
\newpage
\newpage

\end{document}